\documentclass[twocolumn]{aastex62}

\usepackage{enumitem}
\usepackage{amsmath}
\usepackage{units}
\usepackage{apjfonts}
\usepackage{CJK}

\newcommand{\MESA}{{\tt MESA}}

\newcommand{\Ye}{\ensuremath{Y_{\rm e}}}

\newcommand{\dif}{\ensuremath{\mathrm{d}}}
\newcommand{\Dif}{\ensuremath{\mathrm{D}}}

\newcommand{\ddt}[1]{\frac{\partial #1}{\partial t}} 
\newcommand{\DDt}[1]{\frac{\Dif #1}{\Dif t}} 
\newcommand{\ddm}[1]{\frac{\partial #1}{\partial m}} 

\newcommand{\epsgrav}{\ensuremath{\epsilon_{\mathrm{grav}}}} 
\newcommand{\epsnu}{\ensuremath{\epsilon_{\nu}}} 
\newcommand{\epsne}{\ensuremath{\epsilon_{22}}} 
\newcommand{\epsdiff}{\ensuremath{\epsilon_{\mathrm{diff}}}}

\newcommand{\amu}{\ensuremath{m_{\mathrm{p}}}} 

\newcommand{\mue}{\ensuremath{\mu_{\mathrm{e}}}}

\newcommand{\Abar}{\ensuremath{\langle A \rangle}}
\newcommand{\Zbar}{\ensuremath{\langle Z_{\rm i} \rangle}}

\newcommand{\nuclei}[2]{\ensuremath{\mathrm{^{#1}#2}}}

\newcommand{\carbon}[1][12]{\nuclei{#1}{C}}

\newcommand{\oxygen}[1][16]{\nuclei{#1}{O}}

\newcommand{\neon}[1][20]{\nuclei{#1}{Ne}}
\newcommand{\sodium}[1][23]{\nuclei{#1}{Na}}
\newcommand{\magnesium}[1][24]{\nuclei{#1}{Mg}}

\newcommand{\grad}{\vec{\nabla}}
\newcommand{\vecdot}{\cdot}
\renewcommand*{\emph}[1]{\textit{\textbf{#1}}}

\newlength{\apjcolwidth}
\begin{document}
\begin{CJK*}{UTF8}{gbsn}

\author[0000-0002-4791-6724]{Evan B. Bauer}
\affiliation{Kavli Institute for Theoretical Physics, University of California, Santa Barbara, CA 93106, USA}
\correspondingauthor{Evan B. Bauer}
\email{ebauer@kitp.ucsb.edu}

\author[0000-0002-4870-8855]{Josiah Schwab}
\altaffiliation{Hubble Fellow}
\affiliation{Department of Astronomy and Astrophysics, University of California, Santa Cruz, CA 95064, USA}

\author{Lars Bildsten}
\affiliation{Kavli Institute for Theoretical Physics, University of California, Santa Barbara, CA 93106, USA}
\affiliation{Department of Physics, University of California, Santa Barbara, CA 93106, USA}

\author[0000-0002-9156-7461]{Sihao Cheng (程思浩)}
\affiliation{Department of Physics and Astronomy, The Johns Hopkins University, Baltimore, MD 21218, USA}

\received{2020 July 7}
\revised{2020 September 2}
\accepted{2020 September 3}

\title{Multi-Gigayear White Dwarf Cooling Delays from Clustering-Enhanced Gravitational Sedimentation}

\begin{abstract}

  Cooling white dwarfs (WDs) can yield accurate ages when theoretical
  cooling models fully account for the physics of the dense plasma of WD
  interiors. We use \MESA\ to investigate cooling models for 
  a set of massive and ultra-massive WDs (0.9--1.3~$M_\odot$) for which previous
  models fail to match kinematic age indicators based on {\it Gaia} DR2.
  We find that the WDs in this population can be explained as C/O cores
  experiencing unexpectedly rapid $^{22}$Ne sedimentation in the
  strongly liquid interior just prior to crystallization. We propose
  that this rapid sedimentation is due to the formation of solid
  clusters of $^{22}$Ne in the liquid C/O background plasma.  We show
  that these heavier solid clusters sink faster than individual
  $^{22}$Ne ions and enhance the sedimentation heating rate enough to dramatically
  slow WD cooling.  \MESA\ models including our prescription for cluster
  formation and sedimentation experience cooling delays of
  $\approx$4~Gyr on the WD Q branch, alleviating tension between cooling
  ages and kinematic ages.  This same model then predicts cooling delays
  coinciding with crystallization of 6~Gyr or more in lower mass WDs
  (0.6--0.8~$M_\odot$).  Such delays are compatible with, and perhaps
  required by, observations of WD populations in the local 100 pc WD
  sample and the open cluster NGC 6791.  These results motivate new 
  investigations of the physics of strongly coupled C/O/Ne plasma
  mixtures in the strongly liquid state near crystallization and tests
  through comparisons with observed WD cooling.
  
\end{abstract}


\section{Introduction}

White dwarfs (WDs) are stellar embers that, when isolated from
interaction with other stars, cool and fade over Gyr timescales as
they radiate away their residual thermal energy and eventually
crystallize \citep{Fontaine01,AlthausReview}. WD temperatures and
luminosities therefore serve as important age indicators for these
stars and their environments. Early work on WD cooling found that
timescales for WD cooling depend primarily on thermal transport in the
outer layers and the total thermal energy of the ionized plasma in the
WD interior \citep{Mestel52}. Later work showed that as a WD becomes
fainter and its interior cools, it is important to account for other
sources of energy such as the latent heat released by interior phase
transitions \citep{vanHorn68},
chemical separation \citep{Mochkovitch83,Isern1991,Segretain94},
and heavy element sedimentation \citep{Bildsten2001}.

Recent data from the {\it Gaia} mission \citep{GaiaCollaboration2016} 
have greatly enriched the sample of known WDs
\citep{Bergeron2019,Coutu2019,McCleery2020,Kilic2020} and enabled new
tests of WD cooling theory and its applications. In particular, 
\citet{Cheng2019} recently studied a sample of nearby (within 250~pc),
ultra-massive WDs from the
\citet{GentileFusillo2019} catalog, which is based on \textit{Gaia}
DR2 \citep{GaiaCollaboration2018a, GaiaCollaboration2018b}.
Using the transverse velocity as a dynamical age indicator of this
local population, they found evidence for a sub-population of
WDs that appear to be cooling much more slowly than predicted by WD
cooling theory. \cite{Cheng2019} estimated that this extra cooling
delay may be explained by heat released as $^{22}$Ne sediments in the
WD interiors just before they crystallize.
Our aim in this work is to investigate this claim
using detailed evolutionary models of cooling WDs, and describe a
physically motivated modification to the theory of $^{22}$Ne
sedimentation in WD interiors\ that can bring theoretical WD cooling
models into agreement with observations.

The WDs that are the focus of the \cite{Cheng2019} study
occupy a region of the HR diagram known as the WD ``Q~branch.''
This region is named for the DQ WDs showing spectral
signatures of carbon in their atmospheres that are common in this
region of the HR diagram and likely descend from WD mergers
\citep{Dunlap2015,Coutu2019,Koester2019,Cheng2019}, though DA and DB WDs
can also be found on the Q branch. The overdensity of
massive and ultra-massive WDs ($M\approx 0.9-1.3\ M_\odot$) that forms the Q branch
coincides with interior crystallization \citep{Tremblay19}.

\cite{Cheng2019} argue from transverse velocity
data that 5--10\% of ultra-massive WDs experience a cooling
delay of 6--8~Gyr when they reach this region of the HR diagram.
The total amount of energy available to be released by sedimentation of heavy
elements ($^{22}$Ne in particular) in the liquid interior prior to
crystallization is enough to power a cooling delay of this magnitude
\citep{Bildsten2001,Deloye2002,GarciaBerro2008}, but
WD cooling and single-particle diffusion calculations indicate that
these delays will not be realized unless the sedimentation rates are much higher than expected. We pursue
the hypothesis that the necessary enhancement is caused by $^{22}$Ne clustering and the
subsequently more rapid sinking of larger clusters.
  While some previous WD observations have set lower bounds on the
  order of 1~Gyr for the total cooling delays required from
  $^{22}$Ne sedimentation (e.g., $\approx$0.6~$M_\odot$ WDs in NGC~6791,
  \citealt{Bedin08,GarciaBerro2010}),
  the 6--8~Gyr magnitude of the cooling delay for ultra-massive
  WDs on the Q~branch is by far the longest delay required by
  observational evidence. We therefore calibrate our $^{22}$Ne
  clustering model to satisfy the stringent requirements of the
  Q~branch observations before going on to explore consequences for
  other WD populations.

In this work, we investigate detailed WD cooling models that account
for these physical phenomena using the stellar evolution code
Modules for Experiments in Stellar Astrophysics (\MESA,
\citealt{Paxton2011, Paxton2013, Paxton2015, Paxton2018, Paxton2019}).
In Section~\ref{s.models}, we describe how we use \MESA\ to construct
massive and ultra-massive WD models suitable for cooling timescale calculations.
Section~\ref{s.cooling} presents the physics of WD cooling and
describes implementations in \MESA, along with a set of
standard \MESA\ WD cooling models that provide a baseline expectation
for cooling timescales in models that exclude $^{22}$Ne sedimentation.
In Section~\ref{s.single}, we show that the diffusion coefficients for
individual $^{22}$Ne ions are now well constrained theoretically, and the
resulting single-particle sedimentation speeds are too slow for the
rate at which sedimentation energy must be released to explain the
cooling delay for a subset of ultra-massive WDs on the Q branch.
Section~\ref{s.clustering} introduces \MESA\ models implementing  our
proposed solution: an enhancement to sedimentation speeds due to the
formation of heavier $^{22}$Ne clusters in the strongly-coupled liquid
interior just before crystallization sets in. After calibrating our
model to explain the observations of ultra-massive WDs on the Q branch, we
then explore predictions for cooling delays in lower mass
(0.6--0.8~$M_\odot$) WDs.
Appendices~\ref{sec:derivation}--\ref{sec:charge} provide further
details on the physical formalism we employ for heating associated
with mixing and sedimentation in multicomponent plasmas.

\section{Initial White Dwarf Models}
\label{s.models}

To construct massive and ultra-massive C/O and O/Ne WD models, we begin from the input
files of \citet{Lauffer2018} who previously created a suite of 
WD models with $M \approx 1.0-1.3~M_\odot$.  We use \MESA\ version
r10398 and make two important modifications to their approach. First, we use
the nuclear network \texttt{sagb\_NeNa\_MgAl.net}.
Unlike \texttt{co\_burn\_plus.net}
used by \citet{Lauffer2018}, this network includes \sodium[23] which
is produced in carbon burning and is typically the third most abundant
isotope in an O/Ne WD \citep[e.g.,][]{Siess2006}.
Second, we turn off convective overshoot below the carbon burning flame,
preventing the formation of hybrid CO/ONe WDs around
$\approx 1.1~M_\odot$.

We focus on WDs descended from metal-rich progenitors to maximize the
potential effect of sedimentation heating on their cooling.
We run a $M_{\rm ZAMS} = 8.8~M_\odot$ model with initial metallicity
of $Z = 0.035$, producing a $0.98~M_\odot$ C/O WD.
We also run a ${M_{\rm ZAMS} = 10~M_\odot}$ model with metallicity
$Z = 0.035$ that produces an O/Ne WD of mass $1.06~M_\odot$.
  We are particularly interested in studying long cooling delays due
  to sedimentation of $^{22}$Ne, which appears in a WD interior at
  abundances that reflect the initial CNO abundances of
  the WD progenitor star. WDs that experience cooling delays of many
  Gyr from such a mechanism must descend from stars that formed in
  $\alpha$-element rich environments many Gyr ago. Fe abundances
  produced later in galactic chemical evolution are not
  relevant for the production of $^{22}$Ne in WD interiors, so it is likely
  that the most delayed WDs that we study here are related to the high
  [$\alpha/\rm Fe$] population of old stars in the galactic disk
  \citep{Nidever2014,Hayden2015,Mackereth2019,Sharma2020}.

  We also view our ultra-massive C/O WD models as potentially
  representing WD merger products.  The surface abundances of the R
  Coronae Borealis (RCB) stars, which are thought to be the products of He
  WD - C/O WD mergers, show evidence of nucleosynthesis that must
  occur in the aftermath of the merger \citep[e.g.,][]{Jeffery2011}.
  In particular, the extreme enhancement of $^{18}$O observed in the
  RCB stars \citep{Clayton2007} demonstrates the
  occurrence of CNO-cycle hydrogen burning to produce $^{14}{\rm N}$
  and the first steps of the helium-burning sequence
  $^{14}{\rm N}(\alpha, \gamma)^{18}{\rm F}(e^+\nu)^{18}{\rm
    O}(\alpha,\gamma)^{22}{\rm Ne}$ that, if completed, would produce
  $^{22}$Ne.  Calculations in the RCB context \cite[e.g.,][]{Staff2012}
  indicate that mergers could plausibly lead to enhancements in the 
  $^{22}$Ne mass fraction of $\approx 0.05$ in the $\approx 0.1\,M_\odot$
  hot, dense region created around the interface of the merged WDs.
  While locally significant, when distributed across the WD, this amount of $^{22}$Ne would represent an
  enrichment to the $^{22}$Ne mass fraction of 0.01 or less, so
  the merger process appears likely to provide at most a modest enhancement
  to $^{22}$Ne.  Broadly, we view our ultra-massive C/O WD models as
  representing double C/O WD merger products, in which each WD
  individually processed its initial CNO metallicity into $^{22}$Ne,
  the merger process itself produced negligible further enrichment of $^{22}$Ne,
  and so the $^{22}$Ne abundance throughout the bulk of the final
  WD merger product still approximately reflects the initial metallicity with
  which the stars were formed.

The \citet{Lauffer2018} models completely remove the stellar envelope with
strong winds. This choice is not physically motivated, but allows the models
  to skip the challenging and time-consuming evolution through the
  thermally-pulsing asymptotic giant branch (TP-AGB).  As a result,
  the outer layers of the WD that are assembled during the TP-AGB and
  the final H and He layers are not self-consistently generated in our
  models.
To create a simple He atmosphere, we then add $10^{-3}~M_\odot$
of pure He by accretion on the surface before cooling begins.
Subsequent diffusion at the boundary between this outer He layer and the
underlying C quickly establishes a smooth transition region
at the base of the He envelope.

As DB WDs, our models will not always match the spectral
classification of the many DQ and DA WDs found in the Q branch region
of the color-magnitude diagram (CMD), but their cooling properties will be sufficient
for our study here. Ultra-massive WDs on the Q branch are still
  relatively luminous and hot, and so the cooling timescales of DA and
  DB WDs in this regime are very similar \citep{Camisassa2019}. Only
  after cooling below the Q branch do DA and DB WDs develop
  significantly different cooling timescales. Similarly, we expect
  that DQ WDs do not have markedly different cooling properties from
  DB WDs on the Q branch, though we are not aware of any detailed
  studies that model DQ WD cooling timescales specifically.

In order to study the range of WD masses relevant for the Q branch
($0.9-1.3~M_\odot$), we scale the mass of these models using the
\MESA\ control \texttt{relax\_mass\_scale}.
This preserves the composition
profile as a function of fractional mass coordinate.  As a
consequence, our models cannot explore the effects of any
mass-dependent trends in the chemical composition, but this is not
important for our purposes.  This scaling
procedure allows us to easily construct models that are difficult to
produce from single star evolution, such as C/O WDs with
$M \gtrsim 1.05~M_\odot$ created by WD mergers.

\begin{figure}
\begin{center}
\includegraphics[width=\apjcolwidth]{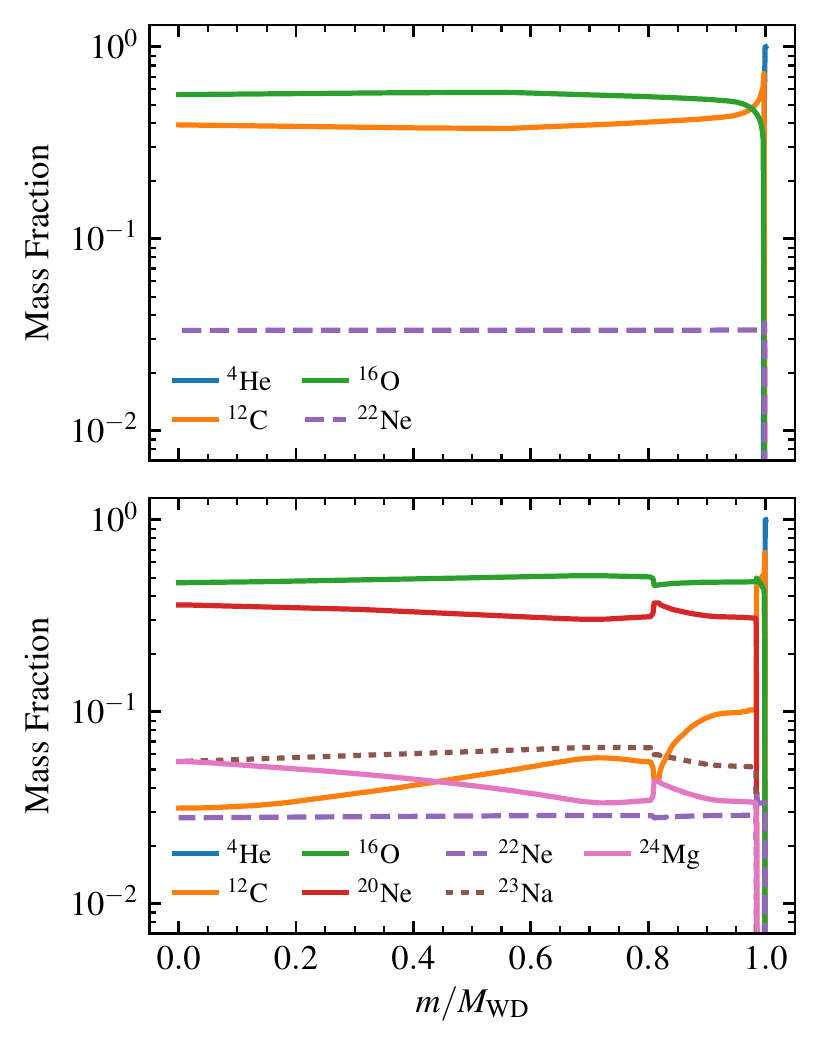}
\caption{Composition profiles of our initial WD models descended from
  $Z=0.035$ progenitors.
  The top panel shows our C/O WD model and the bottom panel
  shows our O/Ne WD model.}
\label{fig:initial_composition}
\end{center}
\end{figure}

  Figure~\ref{fig:initial_composition} shows the initial composition
  profiles of our WD models.  Our C/O WD model can be compared with
  composition profiles from other evolutionary calculations forming massive C/O WDs
  \citep[e.g.,][]{Althaus2010b,Romero2013} and has similar \carbon[12]
  and \oxygen[16] abundances in the core. Our O/Ne WD model can be
  compared with the composition profiles from the ultra-massive WDs of
  \citet{Camisassa2019}.  The mass fractions of \carbon[12],
  \oxygen[16], \neon[20], \sodium[23], and \magnesium[24] are all
  comparable, while our models have more \neon[22] given their higher
  metallicity.  This demonstrates that our more approximate
  evolutionary approach provides composition profiles representative
  of state-of-the-art WD models.

\section{Cooling Models}
\label{s.cooling}

\begin{figure}
\begin{center}
\includegraphics[width=\apjcolwidth]{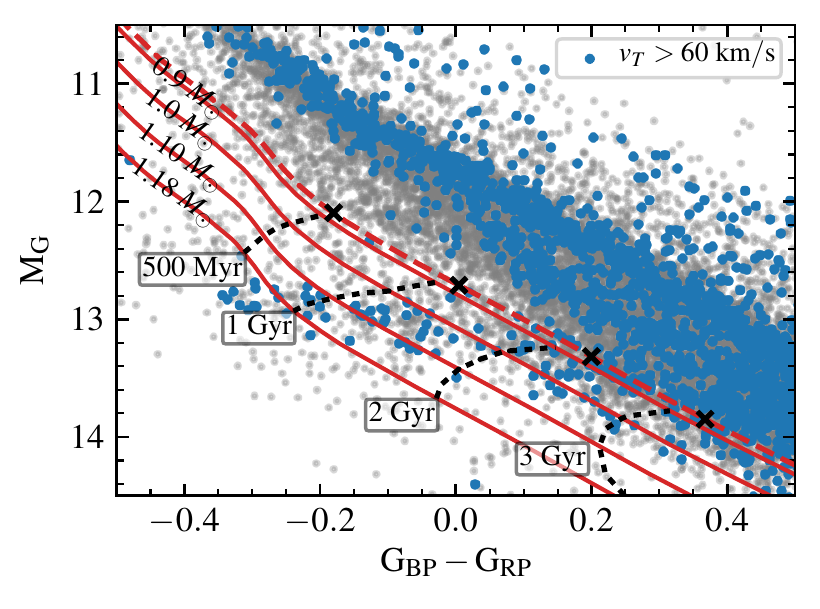}
\caption{C/O WD cooling tracks (red lines) on the {\it Gaia} CMD in the region of
  massive and ultra-massive WDs on the  Q branch (the mostly horizontal feature in the
  data around $\rm M_G \approx 13$). Black dashed curves represent contours
  of constant WD cooling age (not including prior main sequence
  lifetime) for WDs that experience no delay due to heavy element
  sedimentation. The dashed red line shows a Montreal cooling track
  for a 0.9~$M_\odot$ DB WD, and black x's mark ages along that track
  for comparison to the age contours of our \MESA\ WD models.
  Gray points represent WDs within 150~pc according to
  {\it Gaia}, while blue points represent those WDs for which
  {\it Gaia} proper motion measurements also imply velocities transverse to
  our line of sight $v_T > 60\ \rm km/s$.}
\label{fig:standard_cooling}
\end{center}
\end{figure}

The model construction process
yields WDs for which nuclear reactions have ceased. They have
core temperatures of $T_{\rm c} \approx 10^8~\rm K$ and
effective temperatures of $T_{\rm eff} \approx 10^5~\rm K$.
We run WD cooling calculations based on these models using \MESA\
r12115. Figure~\ref{fig:standard_cooling} shows cooling tracks for
massive and ultra-massive C/O WD models on the {\it Gaia} CMD
in the region of Q branch.
The colors for these tracks on the {\it Gaia} CMD are interpolated
based on $\log g$ and $T_{\rm eff}$ from the \MESA\ tracks using the
Montreal synthetic color tables for DB WDs
\citep{Holberg2006,Bergeron2011,Blouin2018}.%
\footnote{\url{http://www.astro.umontreal.ca/~bergeron/CoolingModels/}}
The coverage of these tables places an upper limit on the accessible
mass range for {\it Gaia} colors around 1.2~$M_\odot$. We can run
\MESA\ models more massive than this without difficulty if no colors
are needed, but this mass range will be sufficient for our analysis in
this work.

The WD data and kinematics in
Figure~\ref{fig:standard_cooling} are based on {\it Gaia} DR2
\citep{GaiaCollaboration2018a,GaiaCollaboration2018b} with quality
cuts of \cite{Cheng2019} and limited to a distance of 150~pc.
\cite{Cheng2019} argue that the transverse velocity ($v_T$) data
from {\it Gaia} provide an independent dynamical indicator of the WD ages, and that there is
evidence of a kinematically old population (represented here by
$v_T \gtrsim 60\; \rm km/s$) that must experience a cooling delay of
several Gyr in the region of the Q branch.
This is clearly inconsistent with the much faster cooling rates
exhibited by the tracks in Figure~\ref{fig:standard_cooling}, and this
motivates deeper exploration of the physical processes that may modify
cooling rates in this regime.

Our models have pure He atmospheres, so
we employ a set of atmospheric boundary conditions for DB WDs
based on the \cite{Koester10} WD atmosphere code
(O.~Toloza, 2019, private communication). These tables are important
when WDs cool to $T_{\rm eff} < 15{,}000\ \rm K$ and the conditions
for the cool, dense He at the photosphere run off the \MESA\ opacity
tables as He becomes neutral. These boundary conditions
tabulate pressure and temperature at optical depth $\tau = 25$ where
\MESA\ can attach an interior profile using its gray opacity tables,
similar to what is described for DA WDs in \cite{Paxton2013} based on
the H atmosphere models of \cite{Rohrmann12}.
The tabulated atmospheres for DB WDs are publicly available as a
standard atmosphere option in \MESA\ r12115.

Crystallization in WD interiors releases latent heat that slows WD
cooling by up to $\approx$1~Gyr \citep{vanHorn68,Lamb75,Chabrier00,Fontaine01},
and \cite{Tremblay19} demonstrated that crystallization in their C/O WD
models coincides with the Q branch region on the CMD.
Pile-up during crystallization can explain some of the overdensity of
objects that forms the Q branch, but \cite{Tremblay19} show that this
overdensity is even more strongly peaked than expected from latent
heat release alone. The several Gyr cooling delay inferred by
\cite{Cheng2019} for kinematically old objects (blue points in
Figure~\ref{fig:standard_cooling}) on the Q branch is a delay with
respect to models that already include latent heat release upon
crystallization.

The cooling models in Figure~\ref{fig:standard_cooling}
include the latent heat released by crystallization, but neglect any
heating due to sedimentation of heavy elements, corresponding to the
case of WDs descended from low-metallicity progenitors.
Figure~\ref{fig:standard_cooling} also includes a track for a
0.9~$M_\odot$ DB WD from the Montreal WD cooling models
\citep{Fontaine01}. This track appears slightly offset
from the \MESA\ model of the same mass because the Montreal
models use a 50/50 C/O core composition by mass fraction, while the
ashes of He burning in our \MESA\ models leave behind a core
composition that is closer to 60\% oxygen by mass, resulting in
slightly more compact C/O WD cores. The Montreal models do not include
any heavy element sedimentation in the interior, and the cooling
timescales for the Montreal model agree well with the \MESA\ models
that do not include sedimentation heating.

\subsection{Crystallization}

The onset of crystallization can be quantified using the average
Coulomb coupling parameter of ions in the WD interior plasma:
\begin{equation}
\label{eq:GammaDef}
\Gamma \equiv \frac{\langle Z_{\rm i}^{5/3}\rangle e^2}{a_{\rm e} k_{\rm B}T}~,
\end{equation}
where $a_{\rm e} \equiv (3/4\pi n_{\rm e})^{1/3}$ is the electron
separation and $Z_{\rm i}$ is ion charge.
The plasma is strongly coupled for $\Gamma \geq 1$, and a
classical one-component plasma (OCP) experiences a phase transition from
liquid to solid for $\Gamma = 175$ \citep{PC10}. In a WD interior,
$\Gamma$ increases as the WD cools, and crystallization begins in the
center when $\Gamma$ passes a critical threshold $\Gamma_{\rm crit}$.
For a WD interior composed of a single species of
ions (e.g.\ pure C or pure O), this critical value for crystallization
matches the OCP value ($\Gamma_{\rm crit} = 175$), but for
plasma mixtures the phase curve is more complex
\citep{Medin10,Horowitz10,Blouin2020}.
  For the mixtures dominated by two elements such as C/O or O/Ne
  characterizing WD interiors, the presence of trace impurities such
  as $^{22}$Ne does not appear to have a significant impact on the
  overall phase curve, which can be adequately described by
  calculations for two-component mixtures \citep{Hughto2012}.
For mixtures such as C/O or O/Ne,
we approximate the results of \cite{Medin10} and
\cite{Blouin2020} by adopting a crystallization temperature that is
the critical temperature of a plasma composed purely of the lighter
element of the mixture. That is, crystallization occurs for
\begin{equation}
\label{eq:Gammacrit}
\Gamma = \frac{\langle Z_{\rm i}^{5/3}\rangle e^2}{a_{\rm e} k_{\rm B}T} >
\Gamma_{\rm crit} = 175\frac{\langle Z_{\rm i}^{5/3} \rangle}{Z_{\rm light}^{5/3}}~.
\end{equation}
This approximation agrees with \cite{Medin10} to within about 5\% for
mixtures where the heavier element is not too abundant by number fraction:
$Y_{\rm heavy} \lesssim 0.5$, which is valid for the C/O and O/Ne
dominated interiors of the WD models in this work.
Equation~\eqref{eq:Gammacrit} gives $\Gamma_{\rm crit} \approx 230$ for
C/O mixtures and $\Gamma_{\rm crit} \approx 210$ for O/Ne mixtures.

Our \MESA\ models do not include phase separation of C/O
upon crystallization, which may provide an additional delay of up to
1~Gyr coinciding with crystallization
\citep{Mochkovitch83,Segretain93,Segretain94,Chabrier00,Althaus12}.
However, the magnitude of this delay is much smaller than the several
Gyr delay on the Q branch required by the analysis of
\cite{Cheng2019}, and we do not expect phase separation to change the
location of crystallization on the CMD significantly.
For simplicity, we therefore ignore phase separation in our models
for this work.

\begin{figure}
\begin{center}
\includegraphics[width=\apjcolwidth]{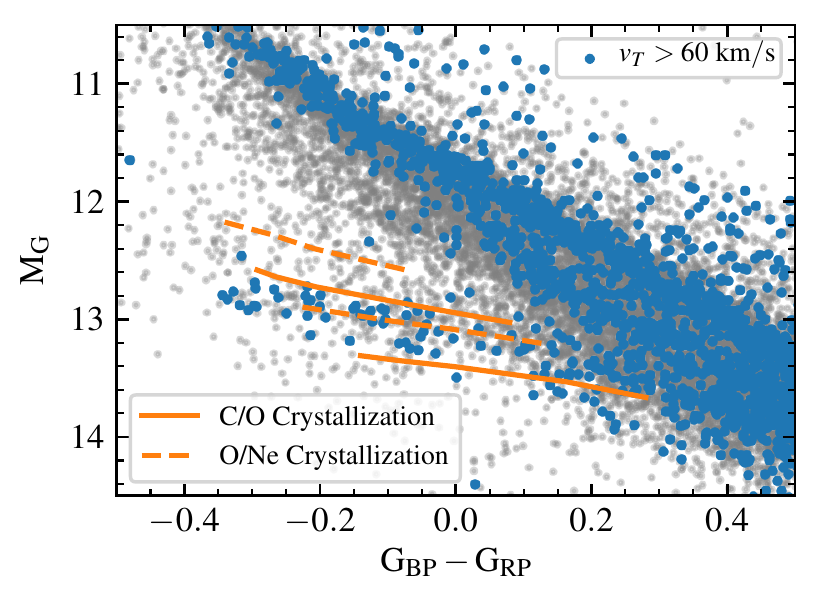}
\caption{Crystallization in ultra-massive O/Ne and C/O WD models compared to the
  {\it Gaia} WD sample within 150~pc. Contours are for cooling
  models of $M=0.9\text{-}1.18\ M_\odot$, with the upper contour
  corresponding to the location where 20\% (by mass) of the interior
  has crystallized and the lower contour where 80\% has crystallized,
  as in \cite{Tremblay19}.}
\label{fig:crystal}
\end{center}
\end{figure}

The \MESA\ treatment of crystallization in WDs employs the \cite{PC10}
equation of state (EOS) and allows $\Gamma_{\rm crit}$ to be set as a
user-defined parameter \citep{Paxton2018}, so our WD models set this
parameter according to Equation~\eqref{eq:Gammacrit}. Due to the
higher average charge of ions in O/Ne WDs, Equations~\eqref{eq:GammaDef}
and~\eqref{eq:Gammacrit} predict that they will crystallize at
interior temperatures that are $\approx$50-60\% higher than C/O WDs of
the same density. Figure~\ref{fig:crystal} shows that this difference
translates into a significant offset on the CMD between O/Ne and C/O
crystallization sequences for ultra-massive WDs. The contours in Figure~\ref{fig:crystal}
rely on a grid of \MESA\ WD cooling models with $M=0.9\text{-}1.18\ M_\odot$.
These models have helium atmospheres and include
crystallization and latent heat release as part of the evolution.
Hydrogen atmospheres change the colors somewhat for any given model,
but the overall location of the crystallization contours does not
change much for different atmosphere compositions.

The location of the Q branch WD overdensity on the CMD is
inconsistent with O/Ne crystallization, but is consistent with C/O
cores even for WDs with mass $M > 1.05~M_\odot$. Furthermore,
the delays due to sedimentation heating that will be discussed in
subsequent sections require that the WD interior remain liquid in order for
diffusion to be active. After crystallization, no further
sedimentation heating is possible, and the cooling cannot be slowed to
form an overdensity of kinematically old WDs on the CMD.
The kinematics and current CMD
positions of WDs in Figure~\ref{fig:crystal} require that most of the
several Gyr cooling delay inferred by \cite{Cheng2019}  occurs after
reaching cooler temperatures than allowed by O/Ne crystallization if
heavy element sedimentation is the cause.
Only the C/O crystallization sequence is consistent with the location
of the Q branch CMD overdensity.

These WDs on the Q branch with $M > 1.05\ M_\odot$ and C/O cores may
be the products of C/O WD mergers that avoid core C ignition
during the merger process. Using velocities of WDs above the Q branch, 
\cite{Cheng2019,Cheng2020} estimate that about 30\% of WDs in this
mass range originate from WD mergers,
in line with recent binary population synthesis predictions of
\cite{Temmink20}. \cite{Kilic2020} reach a similar conclusion for
massive WDs in their sample of spectroscopically confirmed DA WDs
within 100~pc in the SDSS footprint.
Indeed, \cite{Dunlap2015} have also argued that ultra-massive DQ WDs
are likely WD merger products, and \cite{Coutu2019} and
\cite{Cheng2019} recently found that about 45\% of ultra-massive DQ WDs, a
significant fraction of WDs on the Q branch, are kinematically old,
thus indicating a clear relationship between WD mergers and the
cooling delay.

If WD mergers are the only way to produce C/O WDs in
this mass range, Figure~\ref{fig:crystal} clearly suggests that a high
merger fraction among this ultra-massive WD population is needed to produce
the overdensity of WDs on the CMD aligned with C/O crystallization,
which is then further accentuated by the cooling delay that some C/O
WDs on the Q branch experience. There is no apparent CMD overdensity of
WDs corresponding to O/Ne crystallization, which limits the fraction
of WDs in this mass range produced by single-star evolution or any
cooling delay they might experience associated with O/Ne
crystallization. As we are primarily seeking to explain the cooling
delay for the kinematically old WDs (blue points) that sit in the C/O
crystallization regime in Figure~\ref{fig:crystal},
we shall proceed with C/O cooling models for much of the remainder of
this work, even for masses well above 1.0~$M_\odot$.

\section{Single-Particle Sedimentation Heating}
\label{s.single}

Diffusion and sedimentation of ions in WD interiors can rearrange the
charge and mass distribution, releasing heat that slows the rate of WD
cooling \citep{Isern1991,Bildsten2001,Deloye2002,GarciaBerro2008}. 
In this section, we argue that cooling delays comparable
to the several Gyr inferred by \cite{Cheng2019} can be explained by
sedimentation of $^{22}$Ne, but only if it occurs much faster than
predicted by single-particle diffusion calculations.
While the total energy available from this sedimentation is more than
sufficient to provide the observed cooling delay, the required speed
is large compared to expectations for $^{22}$Ne diffusive
sedimentation based on single-particle diffusion
coefficients. Therefore we will consider the possibility of $^{22}$Ne
cluster formation that substantially enhances the sedimentation rate.

\subsection{Sedimentation of Neutron-rich Isotopes}

In degenerate WD interiors, the electric field balances the
gravitational force acting on the ions of average charge-to-mass ratio
\citep{Bildsten2001,Chang2010}:
\begin{equation}
eE = \frac{\langle A \rangle}{\langle Z_{\rm i} \rangle}m_{\rm p} g
\approx 2 m_{\rm p} g~.
\end{equation}
The net gravitational sedimentation force on ions of species $j$ is
therefore
\begin{equation}
F_{g,j} \equiv - A_j m_{\rm p} g + Z_j eE 
= - \left(A_j - Z_j \frac{\langle A \rangle}{\langle Z_{\rm i} \rangle}\right) m_{\rm p} g~.
\end{equation}
The quantity $A_j - Z_j \langle A \rangle / \langle Z_{\rm i} \rangle$
is the number of extra neutrons of an isotope relative
to ions of average charge-to-mass ratio in the background plasma.
This term is large for neutron-rich isotopes, such as $^{22}$Ne found
in C/O WD interiors, and both $^{22}$Ne and $^{23}$Na in O/Ne WD
interiors.

For stars massive enough to burn hydrogen through the CNO cycle, this process converts the primordial CNO abundance into $^{14}$N, which
then undergoes the $\alpha$-capture sequence
$^{14}{\rm N}(\alpha, \gamma)^{18}{\rm F}(e^+\nu)^{18}{\rm O}(\alpha,\gamma)^{22}{\rm Ne}$
during He-burning stages. Therefore most of the
initial metallicity $Z$ of these stars will become $^{22}$Ne
distributed evenly through the interior profile of C/O or O/Ne WDs.
In addition, more massive stars that ignite interior carbon burning on
the AGB before shedding their envelopes to become O/Ne WDs also
produce a significant amount of $^{23}$Na.  This occurs as a result of the
$^{12}$C~+~$^{12}$C reaction that branches to $^{23}$Na + p with $\approx 50\%$ probability.
The overall neutron content of the mixture is enhanced
through subsequent captures of the proton and electron capture or positron emission by those products.
Importantly, these $^{12}$C burning products are metallicity-independent
and so $^{23}$Na is generically present in the interior profile of O/Ne WDs at a mass fraction
of $X_{\rm Na} \approx 0.05$ \citep[e.g.,][]{Siess2006, Siess2007}.

The WD cooling delay caused by a total energy
$\mathcal E$ released at WD luminosity $L_{\rm WD}$ is
${t_{\rm delay} \sim \mathcal E/L_{\rm WD}}$. We will describe these energies in
terms of the potential energy change associated with particles of
species $j$ in the plasma traveling a distance $R_{\rm WD}$ under the
influence of force $F_j$:
$\mathcal E \sim |F_j| R_{\rm WD} X_j M_{\rm WD}/A_j m_{\rm p}$, where
$X_j$ is the mass fraction of element $j$ and $A_j$ is its atomic mass
number. Taking a luminosity of $L_{\rm WD}
\approx 10^{-3}~L_\odot$ as representative of WDs on the Q branch, and
$g\approx 10^{9}~\rm cm\,s^{-2}$ as the typical local gravity in the
interior of a 1~$M_\odot$ WD, we can express the upper limit for the
total time delay as 
\begin{equation}
\label{eq:delay}
\begin{aligned}
t_{\rm max} \approx &10~{\rm Gyr}
\left(\frac{X_j/A_j}{10^{-3}}\right)
\left(\frac{|F_j|}{m_{\rm p} g}\right) \\
&\times
\left(\frac{g}{10^9~\rm cm\,s^{-2}}\right)
\left(\frac{10^{-3}~L_\odot}{L_{\rm WD}}\right)
\left(\frac{R_{\rm WD}}{10^{-2}~R_\odot}\right)
\left(\frac{M_{\rm WD}}{M_\odot}\right)~.
\end{aligned}
\end{equation}
This represents the total available
time delay from a given energy source as it requires all of the
element in question to move through the entire potential of the star.
The maximum time-delay predicted by Equation~\eqref{eq:delay}
for $^{22}$Ne sedimentation is $\approx({\rm 10~Gyr})(Z/0.01)$,
while $^{23}$Na could provide an additional delay up to $\approx$25~Gyr.

Another potential driver of sedimentation and heating in WD interiors
is charge stratification in strongly coupled plasma,
the ion chemical potentials pushing ions of higher charge toward the
center even for ions of average charge-to-mass ratio
\citep{Chang2010,Beznogov2013}.
In Appendix~\ref{sec:charge} we use Equation~\eqref{eq:delay} to
show that additional heating associated with this physics is
negligible compared to sedimentation driven by the weight of
neutron-rich isotopes, so we ignore any additional sedimentation
heating from charge stratification for our models in this paper.

While Equation~\eqref{eq:delay} quantifies the maximum delay available
from sedimentation energy, accurate prediction of the total time delay
achieved from diffusive sedimentation requires detailed models of WD cooling
and interior diffusion \citep{Bildsten2001,Deloye2002,GarciaBerro2008}.

\subsection{Single-Particle Diffusion Coefficients}

The rate at which $^{22}$Ne sedimentation deposits heat in the
interior depends directly on the diffusion velocity of $^{22}$Ne:
$\epsilon_{22} \propto v_{\rm Ne}$ (see Equation~\eqref{eq:epsdiff} and
Appendix~\ref{sec:conceptual}). For trace $^{22}$Ne sinking in the WD
interior, this velocity is
\begin{equation}
\label{eq:vNe}
v_{\rm Ne} = 2 m_{\rm p} g \frac{D_{\rm Ne}}{k_{\rm B} T}~,
\end{equation}
and so $\epsilon_{22} \propto D_{\rm Ne}$, where $D_{\rm Ne}$ is the
single-particle $^{22}$Ne diffusion coefficient. Uncertainties in the
value of this coefficient at strong ion coupling ($\Gamma \gtrsim 10$)
have been a source of uncertainty regarding the total rate of $^{22}$Ne
heating in past studies
\citep{Bildsten2001,Deloye2002,GarciaBerro2008}.

\begin{figure}
\begin{center}
\includegraphics[width=\apjcolwidth]{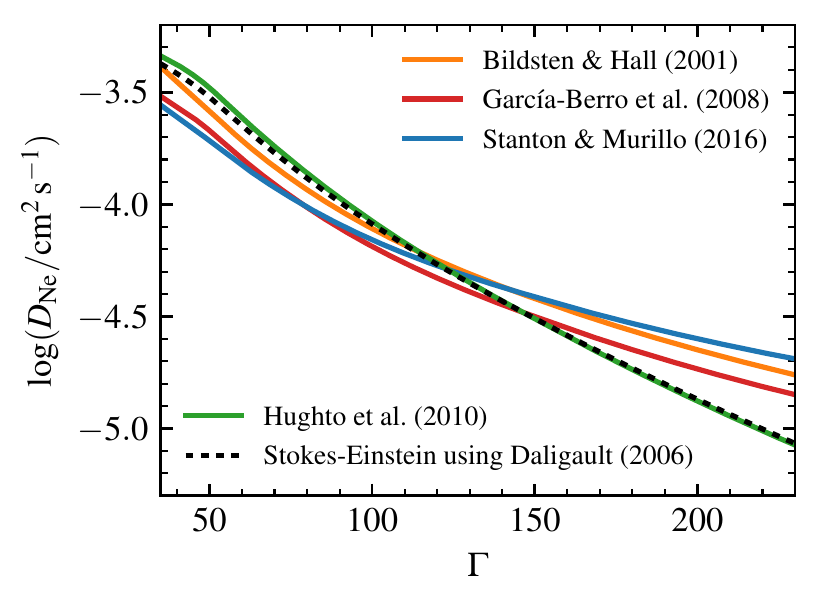}
\caption{Comparison of single-particle diffusion coefficients in the
  interior of a \MESA\ C/O WD model near crystallization.}
\label{fig:coefficients}
\end{center}
\end{figure}

Recent calculations of $D_{\rm Ne}$ in the
strongly coupled regime of C/O WD interiors have reduced the current
uncertainty to no more than a factor of two, and likely substantially
smaller. \cite{Hughto2010} performed molecular dynamics (MD)
simulations of C/O plasma mixtures with trace $^{22}$Ne characteristic
of WD interiors over a plasma coupling range including strongly liquid
conditions ${1 < \Gamma < 244}$.
They calculated $D_{\rm Ne}$ based on these simulations by
measuring the velocity autocorrelation functions for $^{22}$Ne
particles and provided a fitting formula for $D_{\rm Ne}$.
Figure~\ref{fig:coefficients} shows
$D_{\rm Ne}$ using the \cite{Hughto2010} fitting formula in the
interior profile of a \MESA\ C/O WD model for which crystallization
has just begun near the center ($\Gamma_{\rm center} = 237$). The
figure also shows $D_{\rm Ne}$ calculated for the same \MESA\ WD
profile using the methods of two previous studies of $^{22}$Ne
sedimentation in WDs \citep{Bildsten2001,GarciaBerro2008},
demonstrating agreement within a factor of two throughout most of the
WD interior profile.

Figure~\ref{fig:coefficients} also shows $D_{\rm Ne}$ based on the
coefficients calculated by \cite{Stanton2016}, which are the default
diffusion coefficients in \MESA. This formalism calculates diffusion
coefficients based on binary collision integrals between each pair of
species in the plasma. In the limit of a
trace (Ne) sedimenting through a fixed background (C and O), the total
diffusion coefficient can be expressed in terms of a sum over the
binary resistance coefficients $K_{ij}$ used in the \cite{Burgers1969}
diffusion formalism as
\begin{equation}
D_{\rm Ne,SM} = \frac{n_{\rm Ne} k_{\rm B} T}{\sum_i K_{i,\rm Ne}}
\approx \frac{n_{\rm Ne} k_{\rm B} T}{K_{\rm O,Ne} + K_{\rm C,Ne}}~,
\end{equation}
where the sum can be reduced to just the terms for C and O
as these form the dominant background in which most collisions that
mediate diffusion occur.

Finally, \cite{Bildsten2001} noted that in the liquid regime, an
estimate of the single-particle effective radius $R$ yields the
Stokes-Einstein diffusion coefficient:
\begin{equation}
\label{eq:DSE}
D = \frac{k_{\rm B} T}{4 \pi \eta R}~,
\end{equation}
where $\eta$ is the viscosity of the liquid. Using MD simulations of an
OCP \cite{DaligaultOCP} has shown that in the
strongly liquid regime ($\Gamma \gtrsim 30$)
Equation~\eqref{eq:DSE} surprisingly holds even for an individual ion
when the radius is taken to be $R = 0.73 a_{\rm i}$, 
where $a_{\rm i} = (3/4\pi n_{\rm i})^{1/3}$ is the ion spacing.
It is straightforward to generalize this result to estimate a
Stokes-Einstein diffusion coefficient for ions in a plasma
mixture. \cite{DaligaultOCP} gives the following
fit based on MD for the viscosity of the OCP in the strongly liquid regime:
\begin{equation}
\eta = 0.10 m_{\rm i}  n_{\rm i} a_{\rm i}^2 \omega_{\rm p} e^{0.008\Gamma}~,
\end{equation}
where $m_{\rm i}$ is the mass of the ions in the OCP,
and $\omega_{\rm p}$ is the plasma frequency defined below. In order the extend this
to an average over ions in a mixture, we use $\Gamma$ as defined for a
mixture in Equation~\eqref{eq:GammaDef}, along with
$m_{\rm i} = \langle A \rangle m_{\rm p}$ and
\begin{equation}
\omega_{\rm p} = 
\left( \frac{4 \pi n_{\rm i} \langle Z_{\rm i}^2 \rangle e^2}{\langle A \rangle m_{\rm p}} \right)^{1/2}~.
\end{equation}
For a mixture, the ion density is
${n_{\rm i} = \rho/\langle A\rangle m_{\rm p}}$, and the electron
density satisfies ${n_{\rm e} = \langle Z_{\rm i} \rangle n_{\rm i}}$. For an
ion species of charge $Z_j$ different from the background
$\langle Z_{\rm i} \rangle$,
we rescale the effective radius for Stokes-Einstein drift to account
for the different size of its charge-neutral sphere in a background of
fixed electron density $n_{\rm e}$, with the result
\begin{equation}
\label{eq:Rpart}
R_j = 0.73 a_{\rm i} \left( \frac{Z_j}{\langle Z_{\rm i} \rangle} \right)^{1/3}~.
\end{equation}
For $^{22}$Ne in a C/O background
${R_{\rm Ne} = 0.82 a_{\rm i}}$, and the
black dashed curve in Figure~\ref{fig:coefficients} shows
the resulting diffusion coefficient when using this radius along with
Equation~\eqref{eq:DSE}.

Figure~\ref{fig:coefficients} demonstrates that although
\cite{Hughto2010} make no mention of Stokes-Einstein diffusion in
relation to their MD simulations for C/O/Ne mixtures, their fits for
diffusion coefficients closely match the Stokes-Einstein prediction in
the liquid regime when using a radius scaled to the OCP results of
\cite{DaligaultOCP}. The Stokes-Einstein scaling for diffusion
coefficients holds throughout the liquid WD interior, and so we will
continue to make use of the prediction of Equation~\eqref{eq:DSE} in
later sections whenever a coefficient is needed for particles where
the size $R$ can be calculated.

As a baseline for Ne diffusion in liquid WD interiors,
we adopt the coefficient fitting formula of \cite{Hughto2010}. In
practice, we implement this by rescaling the Ne diffusion velocity in
the \MESA\ model by a factor of $D_{\rm Ne,Hughto}/D_{\rm Ne,SM}$,
with a smooth transition from the default diffusion based on
\cite{Stanton2016} over the range ${30 < \Gamma < 50}$, where this
results in a modest enhancement to the diffusion speed (less than a
factor of two).

\subsection{Implementation of Sedimentation in MESA}

The implementation of element diffusion in \MESA\ is described in
Section~3 of \citet{Paxton2018}, including the addition of a
sedimentation heating source term.%
\footnote{The implementation of sedimentation in \MESA\ differs from the
approach taken by \citet{GarciaBerro2008}.  In
Appendix~\ref{sec:conceptual}, we clarify the difference between these
treatments and show that the net energetics are equivalent.}
In \citet{Paxton2018}, this term included only \neon[22] (which is
typically dominant) but here we modify \MESA\ to include heating
associated with the diffusion of all isotopes included in the nuclear
network.  This allows inclusion of \sodium[23] sedimentation which
is of comparable importance in O/Ne WDs.
The generalized heating term in \MESA\ is implemented as a local
entropy source
\begin{equation}
\label{eq:epsdiff}
\epsdiff = 
\sum_{\rm ions} \left(A_j m_{\rm p} \vec g + Z_j e \vec E \right)
\cdot \frac{X_j \vec v_{{\rm diff},j}}{A_j m_{\rm p}}~,
\end{equation}
where $X_j$ is the mass fraction of ion species $j$, and
$\vec v_{{\rm diff},j}$ is the local diffusion velocity found by the
diffusion solver along with the electric field $\vec E$. Refer to
Appendix~\ref{sec:unifying_paxton} for the formal justification for
this form of the heating term, which is a slightly simplified version
of Equation~\eqref{eq:epsdiff-expanded}. For most C/O WDs,
Equation~\eqref{eq:epsdiff} is dominated by the \neon[22] contribution
and reduces to the form found in Equation~(16) of \cite{Paxton2018}.
A more complete physical description of the heating associated with
chemical transport in multicomponent plasmas is given in
Appendices~\ref{sec:derivation} and~\ref{sec:conceptual}.

\begin{figure}
\begin{center}
\includegraphics[width=\apjcolwidth]{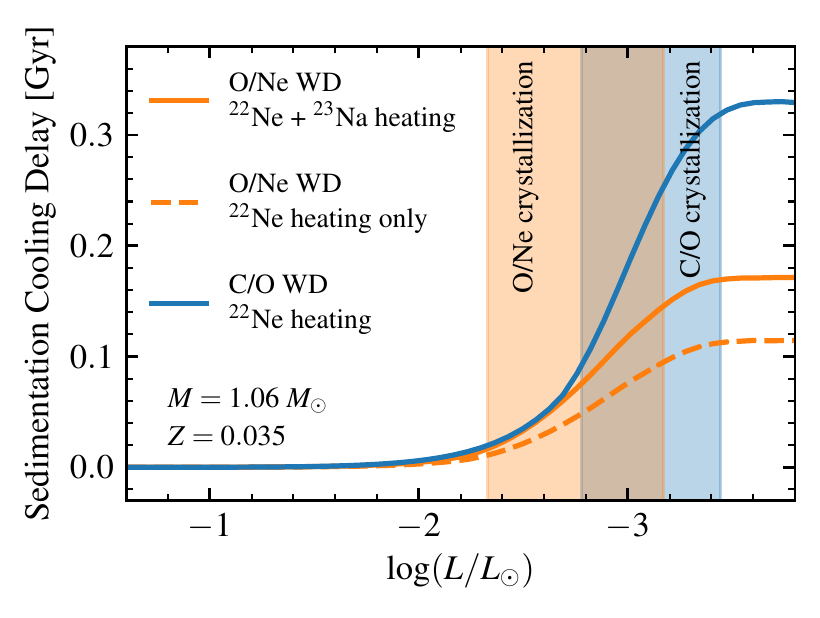}
\caption{
  Accumulated cooling delays due to single-particle sedimentation
  heating relative to models that include no sedimentation heating for
  1.06~$M_\odot$ WD models descended from metal-rich progenitors
  ($Z=0.035$).
  The orange curves show cooling delays for WDs with O/Ne cores, while
  the blue curve shows the cooling delay for a C/O WD.
  The shaded regions indicate the luminosity range over which the
  model interiors pass from 10\% crystallized to 90\% crystallized for
  O/Ne (orange) and C/O (blue) WD cores.
}
\label{fig:age_diff_baseline}
\end{center}
\end{figure}

Figure~\ref{fig:age_diff_baseline} shows cooling delays due to this single-particle
sedimentation heating for 1.06~$M_\odot$ \MESA\ WD models (both O/Ne
and C/O). With the diffusion velocities calculated for single
particles, the neutron-rich isotopes sediment toward the center very
slowly, releasing only a small fraction of the potential heating
available according to Equation~\eqref{eq:delay} before
crystallization sets in and halts sedimentation.
Orange curves in Figure~\ref{fig:age_diff_baseline} show two
cases for the same O/Ne WD model: one for which we include all
sedimentation heating according to Equation~\eqref{eq:epsdiff} (most
importantly $^{22}$Ne and $^{23}$Na), and one for which we only
include the $^{22}$Ne sedimentation heating. While the inclusion of
$^{23}$Na heating does enhance the overall cooling delay for the O/Ne
model, it still provides a much shorter delay than the C/O model of
the same mass experiences from $^{22}$Ne alone (blue curve) due to faster diffusion
of heavy elements in a C/O background than an O/Ne background. For the
C/O case, crystallization (indicated by the shaded regions) also
occurs later, prolonging the sedimentation period and further enhancing
the cooling delay.
The C/O WD model uses the $^{22}$Ne diffusion coefficients based on
\cite{Hughto2010}, while the O/Ne model employs the default \MESA\
diffusion coefficients of \cite{Stanton2016} because we have no MD
results for this case. Even though Figure~\ref{fig:coefficients}
indicates that the \cite{Stanton2016} coefficients are likely too high
in the strongly liquid regime, these O/Ne models still have
significantly shorter cooling delays than the C/O model. It would also
be possible to compute the Stokes-Einstein diffusion coefficients of
Equations~\eqref{eq:DSE}--\eqref{eq:Rpart} for the O/Ne case, but this
would only result in slower diffusion and a smaller cooling delay.

\begin{figure}
\begin{center}
\includegraphics[width=\apjcolwidth]{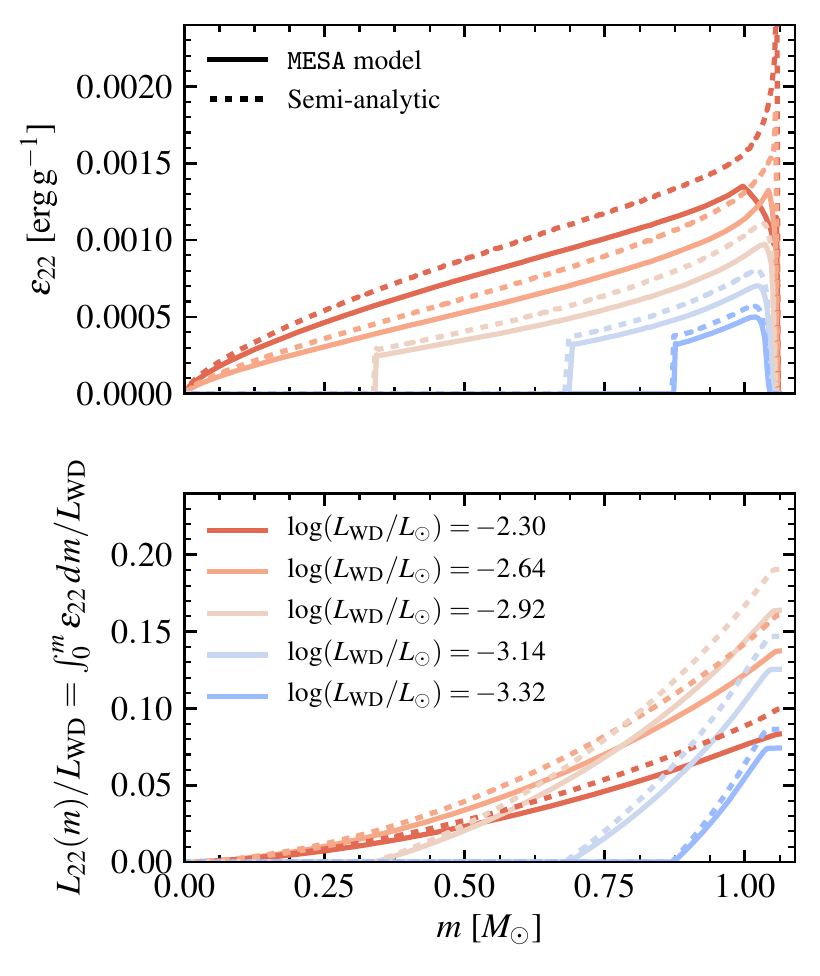}
\caption{$^{22}$Ne heating and integrated luminosity profiles for a
  1.06~$M_\odot$ C/O WD descended from a $Z=0.035$ progenitor. The
  dashed curves show the semi-analytic heating estimate described in
  the text, while the solid curves are the heating profiles from the
  \MESA\ model including the full diffusion solution.
}
\label{fig:heating}
\end{center}
\end{figure}

None of these cases using
single-particle diffusion coefficients experience diffusion that is
fast enough to provide heating adequate to create the several Gyr
cooling delay required by \cite{Cheng2019}.
  Figure~\ref{fig:heating} shows interior heating profiles from
  single-particle $^{22}$Ne sedimentation in the interior of our
  1.06~$M_\odot$ C/O WD model. In order to verify that our heating term in
  \MESA\ matches expectations, we compare to a semi-analytic estimate
  that relies only on the basic thermodynamic structure of the \MESA\
  model ($\Gamma$, $T$, ${X_i}$), approximating
  Equation~\eqref{eq:epsdiff} as
  \begin{equation}
    \epsilon_{22} = 2 m_{\rm p} g \frac{X_{\rm Ne} v_{\rm Ne}}{22 m_{\rm p}}~,
  \end{equation}
  with $v_{\rm Ne}$ given by Equation~\eqref{eq:vNe} and the diffusion
  coefficient $D_{\rm Ne}$ given by the fit of \cite{Hughto2010}. The
  total integrated $^{22}$Ne sedimentation luminosity never exceeds
  20\% of the total WD luminosity, and so this model with
  single-particle sedimentation speeds cannot significantly slow the
  WD cooling to achieve a multi-Gyr delay.

\section{Clustering}
\label{s.clustering}

While the energy available from $^{22}$Ne sedimentation is clearly
adequate to provide the necessary cooling delay observed for Q branch
WDs, this sedimentation must occur significantly faster than predicted
by single-particle diffusion coefficient calculations. Motivated by this as
well as the fact that the cooling delay clearly coincides with
approaching interior crystallization, we hypothesize that clusters of
$^{22}$Ne form in the strongly liquid regime near crystallization.
In this regime, neighboring particles become associated with
each other for many plasma oscillation timescales \citep{Donko2002},
and heavy elements may therefore form clusters that behave as
larger particles even as the background remains liquid,
enhancing the rate of diffusion \citep{Medin2011}.
These clusters will, as we show, sink faster.

The pressure gradient responsible for maintaining hydrostatic
equilibrium in the degenerate WD interior depends primarily
on the electron density $n_{\rm e}$, so we demand that $n_{\rm e}$
remain constant for local variations in ion composition. In this case,
a cluster of particles composed purely of species $j$ will have a
Coulomb coupling parameter of
\begin{equation}
\label{eq:Gammaj}
\Gamma_j = \frac{Z_j^{5/3}}{\langle Z_{\rm i}^{5/3} \rangle} \Gamma~.
\end{equation}
For Ne in a C/O background, this gives
$\Gamma_{\rm Ne} \approx 1.8 \Gamma$. Hence, a group of Ne particles
is much more strongly coupled than the surrounding ions, and may begin
to form solid clusters when the background C/O reaches strongly liquid
conditions. Even if only a fixed cluster for a finite time,
a solid cluster will diffuse and sediment through the
still liquid background as a heavier single particle.
For specific conditions, we expect clusters to have a distribution of
sizes. We use $\langle N \rangle$ to represent the number of Ne atoms
in an average-size cluster. Cluster nucleation has been studied for a
liquid OCP near crystallization \citep{DaligaultNucleation,Cooper2008},
but it is not immediately clear how to
extend this work to mixtures of different particle charges, so we
leave $\langle N \rangle$ as a free parameter for our exploratory
modeling in this work.
Maintaining the same electron density $n_{\rm e}$ as the surrounding
plasma, the radius for a cluster containing $\langle N \rangle$
particles is
\begin{equation}
\label{eq:Rcl}
R_{{\rm cl},j} = a_{\rm i} \left(\frac{\langle N \rangle Z_j}{\langle
    Z_{\rm i} \rangle} \right)^{1/3}~.
\end{equation}
The total downward sedimentation force on a cluster of $\langle N
\rangle$ $^{22}$Ne particles is $2 m_{\rm p} g \langle N \rangle$,
so Stokes-Einstein drift in the liquid regime predicts that the
velocity of cluster sedimentation is
\begin{equation}
v_{\rm cl} = 2 m_{\rm p} g \langle N \rangle \frac{D_{\rm cl}}{k_{\rm B} T}~.
\end{equation}
The diffusion coefficient $D_{\rm cl}$ can be obtained from
Equation~\eqref{eq:DSE} using the radius of Equation~\eqref{eq:Rcl}.
Since we have shown that individual ions diffuse as
spheres of radius given by Equation~\eqref{eq:Rpart} following
Stokes-Einstein drift, we express the cluster diffusion velocity
in terms of a simple rescaling of the individual particle velocities
to larger, heavier clusters diffusing in a background plasma of the
same viscosity. By comparison with Equation~\eqref{eq:vNe},
we can write the diffusion velocity of a cluster of Ne particles in
terms of the diffusion velocity for individual Ne ions as
\begin{equation}
\label{eq:vcl}
v_{\rm cl} = 0.73 \langle N \rangle^{2/3} v_{\rm Ne}~.
\end{equation}
Note that this expression assumes that the diffusion velocity is
dominated by drift due to external forces (gravity and the electric
field) and ignores ``ordinary diffusion'' due to composition
gradients. This is justified in strongly degenerate WD interiors
because the scale height relevant for ion composition gradients is
much smaller than the pressure scale height:
$k_{\rm B} T/m_{\rm p} g \ll P/m_{\rm p} g$. Therefore the composition
gradients driving ordinary diffusion are only significant for sharp
concentrations of elements (e.g.\ pile-up at the center) over a much
smaller scale than gravitational sedimentation of heavy particles
through the bulk of the interior, and thus ordinary diffusion can
be neglected for the bulk heating that we are studying here.

According to Equation~\eqref{eq:vcl}, clusters of $\langle N
\rangle = 100$ particles correspond to a factor of 16 enhancement to
the rate of diffusion (and corresponding sedimentation heating).
We hypothesize that these Ne clusters will form and enhance diffusion
only in strongly liquid conditions past some critical threshold of
$\Gamma_{\rm Ne}^{\rm crit}$, which is a free parameter in our models.
On the other hand, no enhancement to diffusion will occur for WDs that
have not yet cooled enough to approach strongly liquid interiors and
crystallization. The only two free parameters for this theory are
$\langle N \rangle$ and $\Gamma_{\rm Ne}^{\rm crit}$.
The enhancement to diffusion
given by Equation~\eqref{eq:vcl} is active for an interior region from
the time it passes the threshold for cluster formation until it
crystallizes, transitioning into the solid state and freezing out
diffusion.

\subsection{MESA Models Including Cluster Diffusion}
\label{s.MESAcl}

\begin{figure}
\begin{center}
\includegraphics[width=\apjcolwidth]{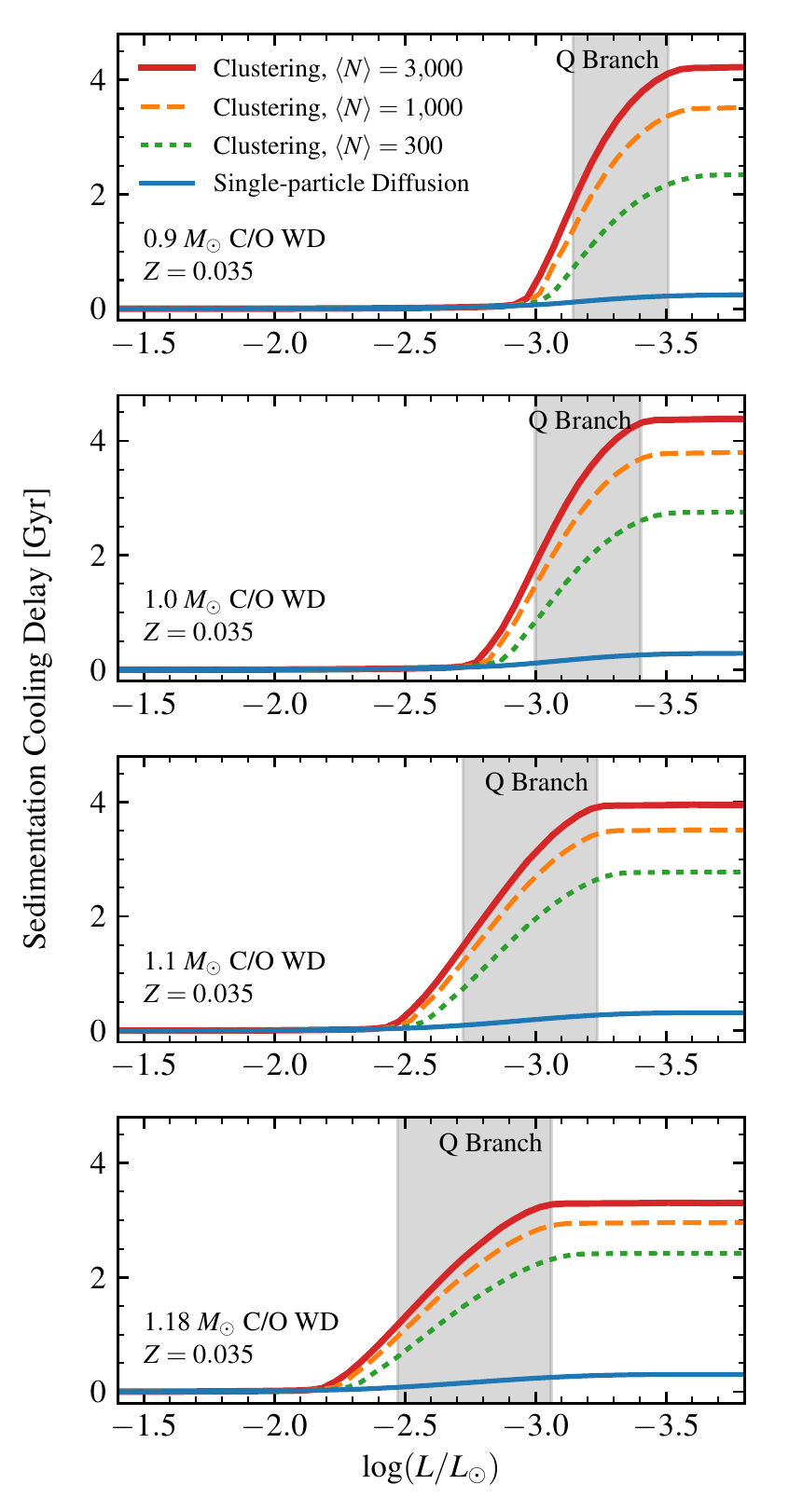}
\caption{
  Accumulated cooling delay due to sedimentation heating relative to
  models that include no sedimentation heating. Blue curves show the
  cooling delay for models where $^{22}$Ne sediments as single particles
  as in Figure~\ref{fig:age_diff_baseline}, while other curves show
  the delay for models in which sedimentation occurs via $^{22}$Ne clusters
  according to Equation~\eqref{eq:vcl} with
  $\Gamma_{\rm Ne}^{\rm crit} = 300$ and various values of $\langle N \rangle$.
  The gray shaded region labeled ``Q~branch'' shows the
  luminosity range where interior crystallization occurs,
  corresponding to the solid contours bracketing the Q branch in
  Figure~\ref{fig:crystal}.
}
\label{fig:age_diff_cluster}
\end{center}
\end{figure}

\begin{figure}
\begin{center}
\includegraphics[width=\apjcolwidth]{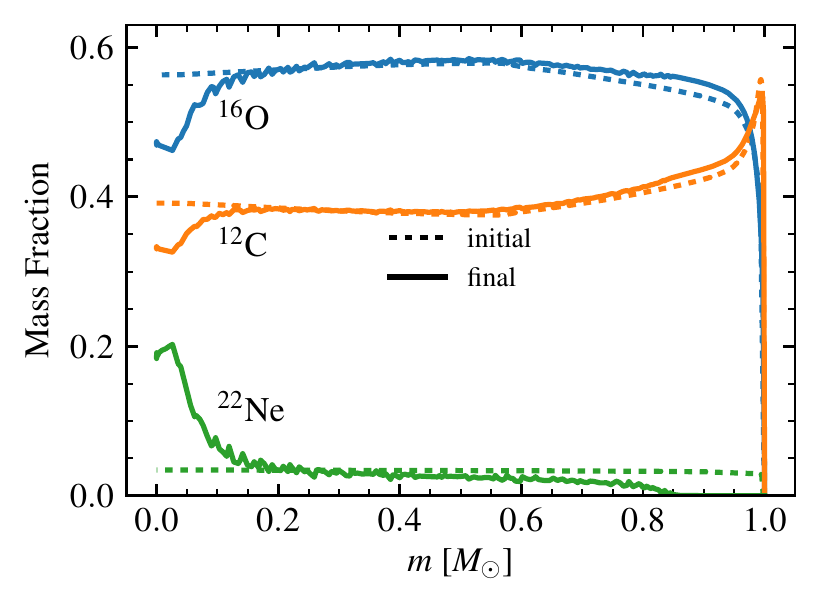}
\caption{
  Interior composition profiles before and after sedimentation for
  the 1.0~$M_\odot$ C/O WD model from
  Figure~\ref{fig:age_diff_cluster} with diffusion of
  $\langle N \rangle = 3{,}000$ clusters.
}
\label{fig:comp}
\end{center}
\end{figure}

We now show that clusters of size $\langle N \rangle \gtrsim 1{,}000$ are
required for an enhancement to diffusion that is sufficient to provide a
cooling delay on the order of several Gyr. Figure~\ref{fig:age_diff_cluster}
shows cooling delays for \MESA\ C/O WD models that implement the
enhancement to diffusion according to Equation~\eqref{eq:vcl} with
$\langle N \rangle = 300\text{--}3{,}000$ and the onset of clustering at
$\Gamma_{\rm Ne}^{\rm crit} = 300$ ($\Gamma \approx 170$).
These WD models are descended from metal-rich progenitors ($Z=0.035$),
so their interiors are rich in $^{22}$Ne, and their cooling delays
represent the longest delays that C/O WDs can achieve from
sedimentation.
Figure~\ref{fig:comp} shows how the $^{22}$Ne profile is rearranged in
the interior over the course of cluster sedimentation.
Models with $\langle N \rangle > 3{,}000$ tend to achieve
similar cooling delays to the  $\langle N \rangle = 3{,}000$ case,
with $^{22}$Ne quickly sedimenting inward as
soon as local conditions pass the threshold for cluster formation. The
overall cooling delay is not particularly sensitive to
$\langle N \rangle$ beyond the value of $3{,}000$ needed to speed
heating enough to achieve this several Gyr delay.
It is possible that once conditions pass the threshold for cluster
formation, $^{22}$Ne simply ``rains'' toward the center in clusters
that continually grow as they encounter other Ne particles while sinking.

\begin{figure}
\begin{center}
\includegraphics[width=\apjcolwidth]{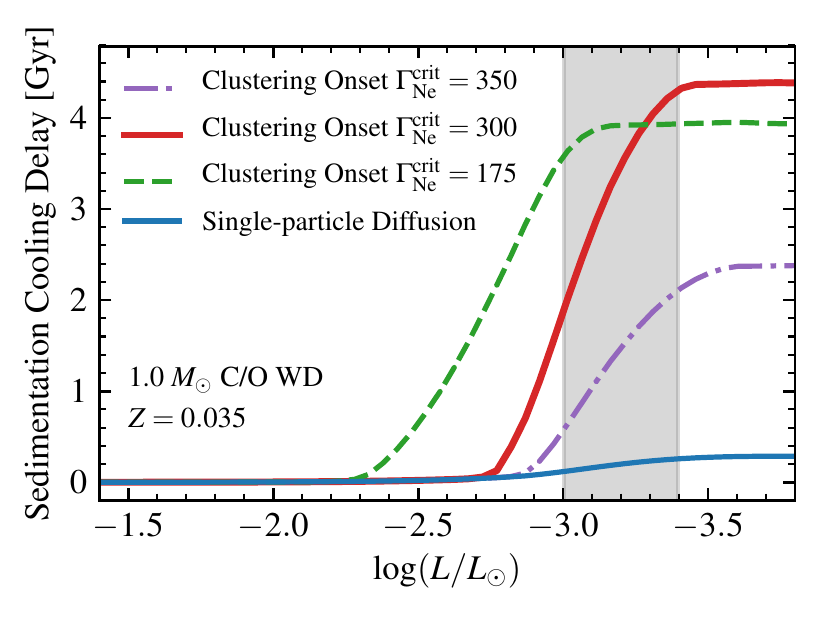}
\caption{
  Comparison of cooling delays for different values of $\Gamma_{\rm
  Ne}$ at which clustering begins to occur in a 1.0~$M_\odot$ C/O WD.
  ${\langle N \rangle = 3{,}000}$ for each of the models including
  clustering here.
}
\label{fig:delay_comparison}
\end{center}
\end{figure}

On the other hand, the delays in these models are quite sensitive to
the value of $\Gamma_{\rm Ne}^{\rm crit}$ where the onset of clustering
occurs. Figure~\ref{fig:delay_comparison} shows a comparison of
cooling delays for a WD model in which only this parameter is varied.
For models in which clustering occurs earlier
than $\Gamma_{\rm Ne}^{\rm crit} = 300$, the cooling delay
mostly accumulates prior to the point where the WD has cooled enough
to reach the Q branch, and the overall magnitude of the delay is
smaller because the sedimentation heating is released while the WD
radiates it away at a higher luminosity. We therefore rule out the
onset of significant clustering for $\Gamma_{\rm Ne} \lesssim 300$ as
inconsistent with the observed CMD location and overall magnitude of
the cooling delay inferred by \cite{Cheng2019}.
For models where $\Gamma_{\rm Ne}^{\rm crit}$ is much larger than 300,
the extent of the range where clustering can operate is limited before
crystallization halts diffusion at
$\Gamma_{\rm Ne} \approx 1.8\times 230 \approx 410$, so we find that
the total delay is maximized around $\Gamma_{\rm Ne}^{\rm crit} = 300$
in our models.
We note however that our WD cooling models do not include any fluid
instabilities that might be triggered in the interior by the
rearranged $^{22}$Ne profile during intermediate stages of
sedimentation. When enhanced $^{22}$Ne sedimentation begins near the
center, it may leave behind a region depleted of $^{22}$Ne that sits beneath
$^{22}$Ne-rich material. The resulting molecular weight gradient could
drive dynamical mixing (e.g., \citealt{Mochkovitch83,Brooks2017}) that further
enhances the net transport of $^{22}$Ne toward the center.
It is currently difficult to account for heating associated with this
dynamical mixing in \MESA\ WD models, so we leave exploration of these
details for future work. This may lead to even longer delays than our
models currently exhibit, particularly for
$\Gamma_{\rm Ne}^{\rm crit} > 300$ where it would extend the region of
the star where enhanced heating can operate to encompass a much larger
portion of the WD core.

\begin{figure}
\begin{center}
\includegraphics[width=\apjcolwidth]{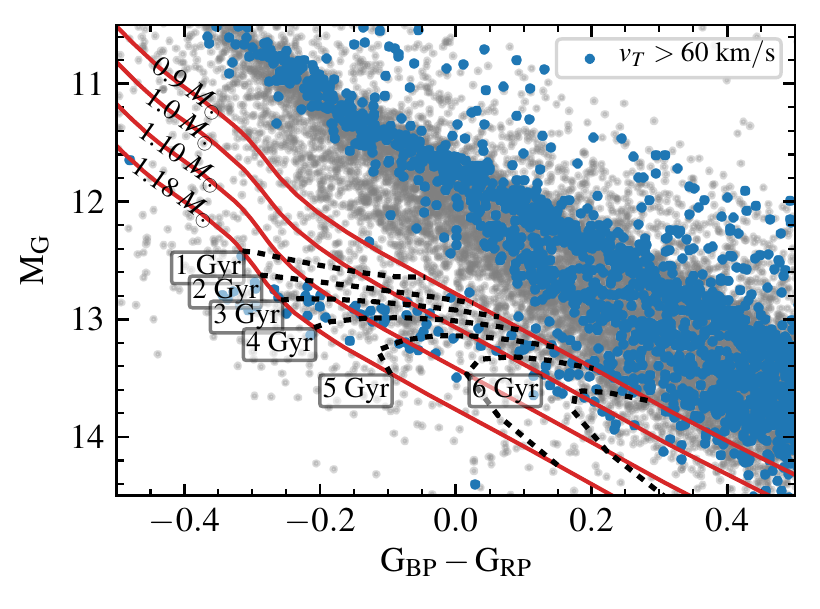}
\caption{
  C/O WD cooling tracks on the {\it Gaia} CMD in the region of the Q branch.
  Black dashed curves represent contours of constant WD cooling age
  (not including prior main sequence lifetime) for WD models including
  $^{22}$Ne sedimentation heating enhanced by clustering for
  $\Gamma_{\rm Ne} > 300$.
}
\label{fig:cluster_cmd}
\end{center}
\end{figure}

Figure~\ref{fig:cluster_cmd} shows cooling tracks and contours of
constant cooling age on the {\it Gaia} CMD for C/O WD models that
include sedimentation heating with diffusion enhanced according to
Equation~\eqref{eq:vcl} with $\langle N \rangle = 3{,}000$ and the
onset of clustering at $\Gamma_{\rm Ne}^{\rm crit} = 300$. The
slowdown in the cooling rate is evident in the higher density of these
contours in the region of the Q branch. For the Q branch to host
enough fast-moving WDs ($v_{\rm T} > 60\ \rm km/s$), \cite{Cheng2019}
calculated that the delayed population should experience an 8~Gyr
delay even after accounting for latent heat and merger delays.
This value may be reduced to $\approx$6~Gyr for a younger thick disk
age (e.g., \citealt{Mackereth2019}) or a higher age--velocity-dispersion 
relation, or if a quickly decreasing star-formation history is
adopted. Our \MESA\ models exhibit a 4~Gyr delay for metallicity
${Z=0.35}$ progenitors without considering phase separation.
Thus our models provide most of the cooling delay needed to explain
velocity observations, though a small tension between these models and
observation still exists.
  \cite{Cheng2019} used O/Ne models for inferences of the cooling
  delay when $M \geq 1.1\ M_\odot$ on the Q~branch, so the C/O core
  compositions for our ultra-massive models would imply an additional
  delay of 1--2~Gyr due to later crystallization and phase separation.
  DA and DQ WDs may also exhibit small differences from our DB
  models in sedimentation cooling delays due to minor variations in
  the $L$-$T_{\rm core}$ relation.
The remaining 2--4~Gyr of cooling delay is likely explained by some
combination of these smaller effects or other dynamical mixing that we
have not considered in our models here.

\subsection{Predictions for Lower Mass WDs}

With the free parameters of our clustering model
tuned to match the observations of ultra-massive WDs on the Q branch,
this model makes predictions for the cooling rates of less
massive WDs.  The WD models in this section are constructed using the
\MESA\ test case \texttt{make\_co\_wd} in version r10398.  We
vary initial mass and metallicity to produce WDs of the desired mass,
and save the WD model when it cools to a luminosity of
10~$L_\odot$. For initial $Z=0.02$, an initial mass of 3.25~$M_\odot$
produces a 0.6~$M_\odot$ WD, and 4.25~$M_\odot$ produces a
0.8~$M_\odot$ WD. For initial $Z=0.04$, these initial masses are
3.1~$M_\odot$ and 4.2~$M_\odot$.
The AGB winds in these models leave some hydrogen in the
atmosphere, so when diffusion is turned on,
the models quickly become DA WDs. Our WD cooling
calculations for these models use \MESA\ r12115 as in previous sections.

\begin{figure}
\begin{center}
\includegraphics[width=\apjcolwidth]{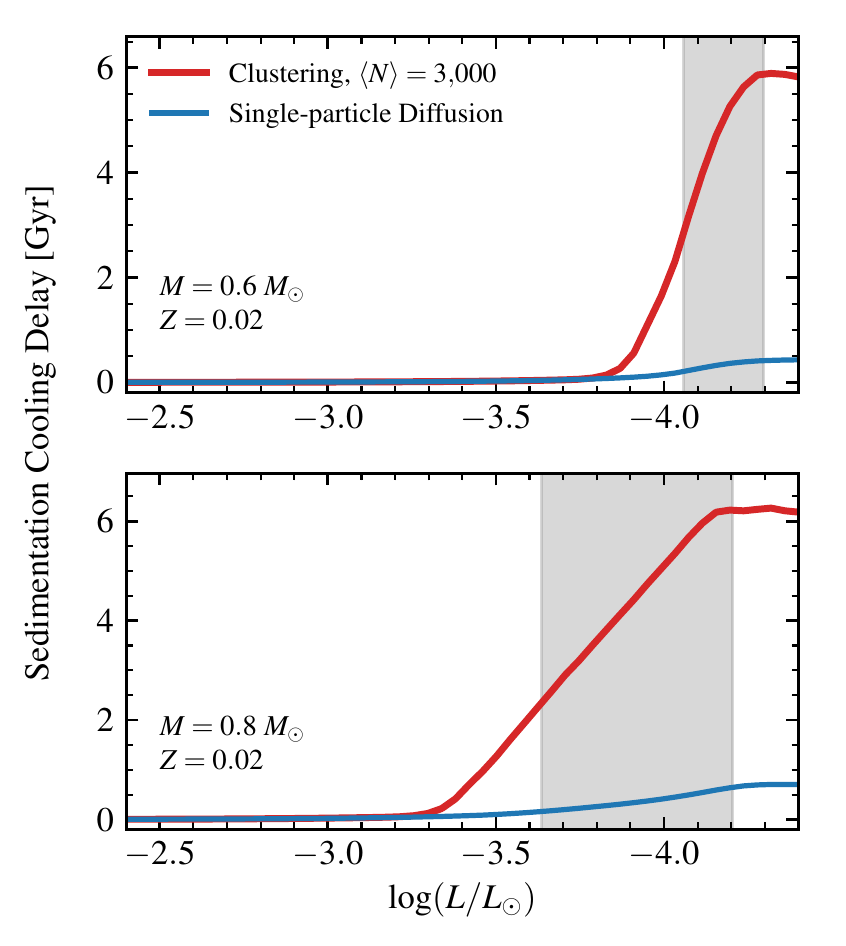}
\caption{
  Cooling delay due to sedimentation heating relative to models that
  include no sedimentation heating. Gray shaded regions indicate the
  luminosity range over which crystallization is occurring in the
  interior.
}
\label{fig:delay_lowM}
\end{center}
\end{figure}

Figure~\ref{fig:delay_lowM} shows the cooling delays for
0.6~$M_\odot$ and 0.8~$M_\odot$ WDs descended from progenitor stars
with metallicity ${Z=0.02}$. Even though this lower metallicity leaves
less $^{22}$Ne in the WD interior compared to the models in the
previous section, sedimentation of $^{22}$Ne powers a
longer cooling delay ($\approx$6~Gyr) in these lower mass models
because crystallization occurs at a much lower luminosity.

Clustering may provide a natural explanation for the 
recent observations of \cite{Kilic2020},
who note that the $M$-$T_{\rm eff}$ distribution
of DA WDs warmer than $6{,}000$~K shows an excess of WDs in the mass
range 0.7--0.9~$M_\odot$.  The location of this pile-up
is coincident with the expected location of crystallization.
However, by comparing the data with cooling models that include
latent heat release during crystallization, \cite{Kilic2020} show that
the latent heat alone does not lead to cooling delays
sufficient to explain the observed distribution.

\begin{figure}
\begin{center}
\includegraphics[width=\apjcolwidth]{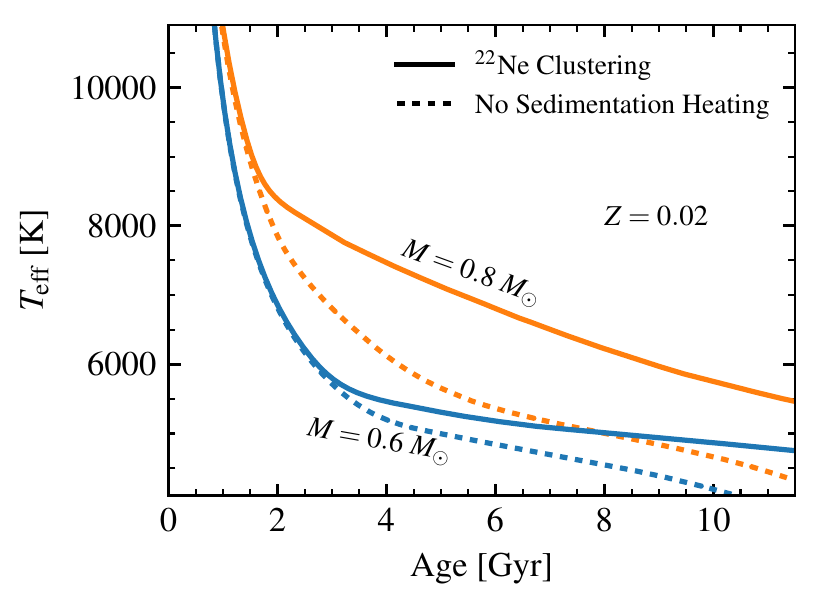}
\caption{
Effective temperature evolution for 0.6 and 0.8~$M_\odot$ models with
and without sedimentation heating from $^{22}$Ne clusters.
}
\label{fig:lowM_cool}
\end{center}
\end{figure}

Figures~\ref{fig:delay_lowM} and~\ref{fig:lowM_cool} show that a
0.8~$M_\odot$ model will experience a significant extra delay due to
$^{22}$Ne clustering coincident with crystallization and consistent
with the temperatures of the excess WDs observed by \cite{Kilic2020}.
On the other hand, clustering will have no effect on the cooling
inferences for 0.6~$M_\odot$ WDs near the peak of the WD mass
distribution in the \cite{Kilic2020} sample because it is limited to
$T_{\rm eff} > 6{,}000\ \rm K$. Clustering and crystallization do not
begin for 0.6~$M_\odot$ WDs until they cool below
$T_{\rm eff} < 6{,}000\ \rm K$.

Due to its high metallicity, $Z\approx 0.04$, the open cluster NGC~6791
was noted by \cite{Deloye2002} as an important environment to probe the impact of
$^{22}$Ne sedimentation on WD cooling. This proved to be the case when the measured
WD luminosity function (WDLF) revealed a peak at $\log(L/L_\odot)\approx-4.0$ that
yielded the faintest WDs in the cluster \citep{Bedin08}.
These WDs should have an age matching the cluster age of
$8.2 \pm 0.3$~Gyr \citep{McKeever19}. However, standard WD cooling models that
do not include $^{22}$Ne sedimentation indicate an age of only 6~Gyr
\citep{Bedin08}. \cite{GarciaBerro2010} and \cite{Althaus10} showed that
including C/O phase separation along with (single-particle)
$^{22}$Ne sedimentation provides a WD cooling delay that is long
enough to alleviate this discrepancy.

\begin{figure}
\begin{center}
\includegraphics[width=\apjcolwidth]{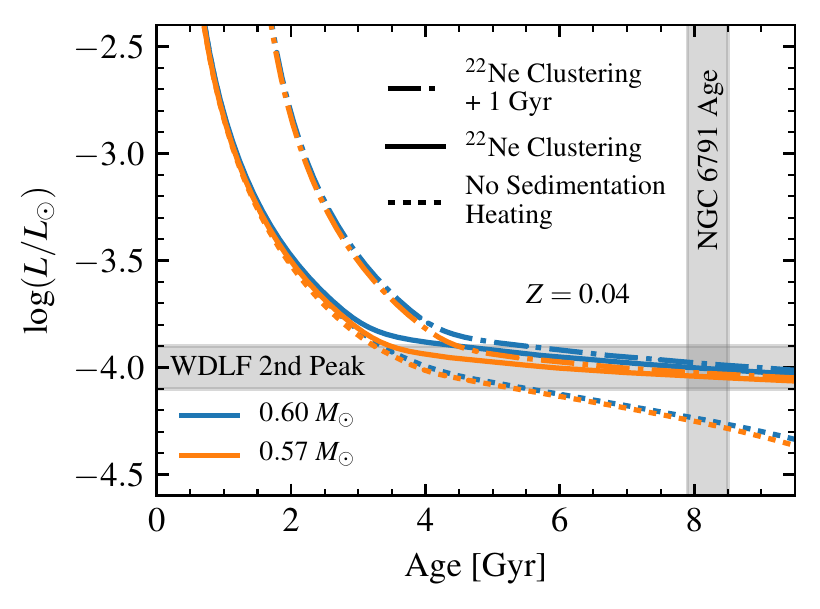}
\caption{
  Luminosity evolution of 0.57 and 0.6~$M_\odot$ WDs compared to
  constraints from observations of the open cluster NGC~6791. Gray
  shaded regions show the cluster age \citep{McKeever19} and faint
  edge of the WD luminosity function \citep{Bedin08}.
}
\label{fig:NGC}
\end{center}
\end{figure}

Figure~\ref{fig:NGC} shows WD cooling models for 0.57 and
0.6~$M_\odot$ WDs with $Z=0.04$, representing
the WD mass range likely to dominate the WDLF peak in NGC~6791.
As previously noted by \cite{Bedin08},
models that do not include sedimentation become too faint
after 6~Gyr.  Our models that include $^{22}$Ne clustering experience
significant cooling delays, but still reach the faint peak of the WDLF
in the cluster age of 8~Gyr.  In \cite{GarciaBerro2010}, the addition
of phase separation and single-particle $^{22}$Ne sedimentation
shifted the brightness of the faint peak in the theoretical WDLF at 8~Gyr by
$\approx -0.5$~mag to bring models into agreement with the observed
WDLF. Comparing the two sets of curves in
Figure~\ref{fig:NGC} at 8~Gyr, our models that include $^{22}$Ne
clustering have $\approx +0.2$~dex higher luminosity than those with
no sedimentation heating. This would shift a theoretical WDLF by
$\approx -0.5$~mag.
  Although our models do not include C/O phase separation during
  crystallization, including an additional delay of $\approx$1~Gyr due
  to phase separation would not significantly change the WDLF
  at the cluster age of 8~Gyr where the evolution is already
  very slow. We show this in Figure~\ref{fig:NGC} by including
  dot-dashed curves with an artificial extra delay of 1~Gyr in
  addition to the $^{22}$Ne clustering. At the cluster age of 8~Gyr,
  the luminosities for these curves are just 5\% higher, or
  $\approx$0.05~mag brighter. Thus, the total shift in the WDLF including
  both $^{22}$Ne clustering and C/O phase separation might be as large
  as $\approx - 0.55$~mag, which would still agree with the
  $\approx -0.5$~mag shift calculated by \cite{GarciaBerro2010} to
  well within the photometric uncertainties ($\approx 0.15$ mag).
Therefore the addition of $^{22}$Ne clustering does not bring
theoretical cooling models into tension with the observations of NGC~6791.

\newpage
\section{Conclusions}

We summarize our main arguments as follows:
\begin{enumerate}[itemsep=2pt,parsep=2pt]
\item A subset (5--10\%) of massive and ultra-massive WDs (0.9--1.3~$M_\odot$)
  experience a cooling delay of several Gyr that is too long to be
  explained by crystallization, phase separation, chemical
  differentiation due to diffusion, or any plausible combination of
  these effects.
\item The evolutionary phase where the cooling delay occurs coincides
  with the regime for a C/O phase transition from liquid to solid in
  the WD core.
\item Heating due to sedimentation of $^{22}$Ne in C/O WD cores can
  provide enough luminosity for a several Gyr cooling delay prior to
  crystallization, but only if sedimentation can proceed much faster
  than predicted by the diffusion coefficients of individual $^{22}$Ne
  ions, which are theoretically well-constrained.
\item We therefore propose that in the strongly liquid regime,
  $^{22}$Ne ions can form into solid clusters of
  $\langle N \rangle \gtrsim 1{,}000$ particles that are heavier and
  sediment toward the center faster according to Stokes-Einstein drift,
  and our \MESA\ models show that this modification creates a
  $\approx$4~Gyr cooling delay, explaining most of the cooling delay
  for fast-moving WDs on the Q branch.
\end{enumerate}
Our simple model for clustering leaves the size of clusters
($\langle N \rangle$) as a free parameter, as well as the value of
Coulomb coupling for Ne at which clusters begin to form
($\Gamma_{\rm Ne}^{\rm crit}$).
While we have not explored a first-principles physical approach
for constraining how these clusters form, it may be that MD modeling
of strongly coupled plasmas (e.g., \citealt{DaligaultOCP,Horowitz10,Hughto2010})
can provide more insight into the physics of this cluster formation.
It may also be the case that solid clusters form with some other
composition than pure Ne. In that case, it would still be
straightforward to reinterpret our sedimentation numbers for solid
clusters composed of a Ne-rich C/O/Ne mixture that sinks relative to
the liquid C/O background.

The subset of ultra-massive WDs that experience this extra cooling delay of
several Gyr due to $^{22}$Ne sedimentation likely descends from
an old, metal-rich population. In particular, the $^{22}$Ne in WD
cores forms from the primordial CNO abundance of their progenitor
stars, so these WDs are likely associated with the high
$[\alpha/\rm Fe]$ sequence of old stars in the galactic disk
population \citep{Nidever2014,Hayden2015,Mackereth2019,Sharma2020}.
Early in galactic history, core-collapse supernovae produced the
CNO necessary to seed the eventual production of
$^{22}$Ne in these old WD cores. The WDs that experience the
longest delays in our model would be those descended from the highest
$\alpha$-abundance progenitors, and the distribution of cooling delays
for this sample of WDs should be correlated with the
$\alpha$-abundance distribution of other old stars within a few
hundred pc of the sun \citep{Castro1997,Pomp2003,Khop2019}.
  Our ultra-massive C/O WD models may represent C/O WD merger
  products, so the final $^{22}$Ne abundance may also be modestly
  enhanced by burning during the merger process (see
  Section~\ref{s.models}).
As most WDs would have lower $^{22}$Ne abundances, they would experience
much less sedimentation heating, which naturally explains why \cite{Cheng2019}
inferred that only a small fraction of WDs experience a substantial
extra cooling delay. We therefore conclude that the sedimentation of
clustered $^{22}$Ne can explain most of the cooling delay on the Q branch,
though some tension between theoretical prediction and observations
may still exist (Section~\ref{s.MESAcl}).

Our model also predicts that lower-mass (0.6--0.9~$M_\odot$) WDs
experience a substantial cooling delay ($\gtrsim$6~Gyr for $Z\geq
0.02$) very late in their evolution. This likely explains the excess
of 0.7--0.9~$M_\odot$ WDs recently observed by \cite{Kilic2020}.
This cooling delay would also have a large impact on the faint edge of
the WD luminosity function in old, metal-rich star clusters.
While high metallicity is rare for old globular clusters
(e.g., \citealt{Hansen2002,Hansen2007,Campos2016}), we have shown that
for the old, metal-rich open cluster NGC~6791, our model produces WD cooling
consistent with the main sequence turnoff age and WD luminosity function.

Although we have shown that WDs with O/Ne cores are unlikely to account
for the cooling anomaly on the Q branch, our clustering model does
predict that O/Ne WDs should also experience a significant cooling
delay coincident with interior crystallization. The $^{23}$Na present
in all O/Ne WD cores should also be susceptible to the clustering
process that we have described for $^{22}$Ne. The charge contrast
between $^{23}$Na ($Z_{\rm i} = 11$) and the O/Ne background is
somewhat smaller than that of $^{22}$Ne to C/O, so clustering would
begin later and possibly only lead to a delay of a few~Gyr, but this
would still slow the cooling rate of O/Ne WDs by a factor of a few.
This only makes it more striking that
Figure~\ref{fig:standard_cooling} shows no feature corresponding to
O/Ne WD crystallization on the CMD, and may imply a stricter
constraint on the fraction of O/Ne WDs among ultra-massive WDs.

When neither $^{22}$Ne and $^{23}$Na are present in a WD interior,
clustering of other elements may play an important role in WD cooling.
For $^{56}$Fe present at mass fraction $X_{\rm Fe} \sim 0.001$,
Equation~\eqref{eq:delay} predicts possible Gyr delays if clustering
can enhance sedimentation sufficiently.
Even in He WD cores that never crystallize, the large charge contrast
of heavier trace elements may put them in the clustering regime
(Equation~\ref{eq:Gammaj}), and the chemical potential for clusters of
elements such as $^{14}$N could drive significant stratification and
heating even without the weight of extra neutrons
(Appendix~\ref{sec:charge}).


\acknowledgments

We are grateful to the anonymous referee for many suggestions that
improved this paper.
We thank Odette Toloza, Boris G{\"a}nsicke, and Detlev Koester for
providing tabulated atmospheric boundary conditions for DB WDs.
We are grateful to Bart Dunlap and JJ Hermes for valuable discussions
about DQ WDs and WD mergers, and to Mike Rich for discussion of
galactic metallicity distributions. We are grateful to Andrew Cumming
and Chuck Horowitz for sharing insights on crystallization in
multi-component plasmas.

This work was supported by the National Science Foundation through
grants PHY-1748958 and ACI-1663688.
This research benefited from interactions that were funded by the
Gordon and Betty Moore Foundation through grant GBMF5076.
This work used the Extreme Science and Engineering Discovery
Environment \citep[XSEDE;][]{XSEDE}, which is supported by National
Science Foundation grant number ACI-1548562, specifically
\textit{comet} at the SDSC through allocation TG-AST180050.
Support for this work was provided by NASA through Hubble Fellowship
grant \# HST-HF2-51382.001-A awarded by the Space Telescope Science
Institute, which is operated by the Association of Universities for
Research in Astronomy, Inc., for NASA, under contract NAS5-26555.
JS is also supported by the A.F. Morrison Fellowship in Lick Observatory.


\software{%
  \texttt{MESA} \citep{Paxton2011, Paxton2013, Paxton2015, Paxton2018, Paxton2019},
  \texttt{ipython/jupyter} \citep{perez_2007_aa,kluyver_2016_aa},
  \texttt{matplotlib} \citep{hunter_2007_aa},
  \texttt{NumPy} \citep{der_walt_2011_aa}, and
  \texttt{Python} from \href{https://www.python.org}{python.org}.
}

\appendix

\section{Derivation of Fundamental Equations}
\label{sec:derivation}

A star is a system with properties that are functions of both space
and time.  Equilibrium thermodynamic systems have homogeneous
properties (i.e., the system has \textit{a} temperature) that do not
vary with time.  Time evolution is typically introduced by thinking
about moving through a sequence of equilibrium states.  The
consideration of spatial variation proceeds by dividing the star into
infinitesimal systems (but not so infinitesimal that thermodynamics
breaks down) and taking the continuum limit.  These infinitesimal
sub-systems are \textit{open} systems that exchange both heat and matter
with their surroundings.

\subsection{The ``First Law'' and Conservation of Energy}
\label{sec:first-law}

The ``first law'' is a statement that encodes energy conservation.
Chapter II of \citet{dGM} demonstrates how to formulate the first law
in a multi-species system where each species is allowed to have a
flow relative to the mean.  We follow their discussion,
but specialize to our problem when it allows simplification.
In order to allow easy comparison with
\citet{Paxton2018}, we use the indices $s$ and $t$ to refer to
particle species.

The hydrodynamic equations come directly from moments of the
collisional Boltzmann equation \citep[see e.g., Chapter 7 of]
[]{Hirschfelder1964}.  When considering the evolution of quantities
that are conserved in collisions (mass, momentum, and energy), sums of
the collision integrals over all species vanish, yielding simpler
equations for their rate of change.

We assume a fully-ionized plasma of $S$ species (electrons plus $S-1$
species of ions).  With no reactions, continuity of each species
yields
\begin{equation}
  \ddt{\rho_s} + \grad \vecdot \left(\rho_s \vec{v}_s\right) = 0~.
  \label{eq:continuity-single}
\end{equation}
The total mass density is
\begin{equation}
  \rho = \sum_{s} \rho_s~,
\end{equation}
and we define the velocity $\vec{v}$ as the barycentric velocity,
\begin{equation}
  \vec{v} \equiv \frac{1}{\rho} \sum_{s} \rho_s \vec{v}_s~.
\end{equation}
We define relative velocities for each species as $\vec{w}_s = \vec{v}_s - \vec{v}$.
Summing the individual continuity equations
(Equation~\ref{eq:continuity-single}) and using the definitions
above gives mass conservation
\begin{equation}
  \ddt{\rho} + \grad \vecdot \left(\rho \vec{v}\right) = 0~.
  \label{eq:continuity-total}
\end{equation}

The inviscid momentum equation is
\begin{equation}
  \ddt{\left(\rho \vec{v}\right)} + \grad \vecdot \left(\rho \vec{v} \otimes \vec{v} \right) = -\grad P +  \sum_{s} \vec{f}_s~.
  \label{eq:momentum-combined}
\end{equation}
where $P$ is the (scalar) pressure and $\otimes$ represents the outer
product.  We assume that
the relative velocities, $\vec{w}_s$, are much less than the thermal
velocities, such that the full pressure tensor is dominated by its scalar
component (the equilibrium pressure).  The force density on species
$s$ is
\begin{equation}
  \vec{f}_s = n_s \vec{F}_s = n_s\left(m_s \vec{g} + q_s \vec{E}\right)~,
  \label{eq:the-force}
\end{equation}
where $n_s$ is the number density, $m_s$ is the mass, $\vec{g}$ is the gravitational acceleration, $q_s$ is the charge, and $\vec{E}$ is the electric field.  Local charge neutrality ($\sum_s n_s q_s = 0$) implies that the total force is simply the gravitational force,
\begin{equation}
  \sum_{s} \vec{f}_s = \rho \vec{g} ~.
  \label{eq:f-sum}
\end{equation}

When in hydrostatic equilibrium, if the ion
contribution to the total pressure is negligible (meaning
$P \approx P_{\rm e}$), the electron equilibrium requires an
approximate balance between the pressure gradient and the electric
force.  This
implies the presence of an
electric field
\begin{equation}
  \vec{E} \approx -\frac{\rho \vec{g}}{\sum_{\rm ions} n_s q_s}~.
  \label{eq:e-field}
\end{equation}

The equation for the specific (per unit mass) internal energy, $u$, is
\begin{equation}
  \ddt{(\rho u)} + \grad \vecdot \left(\rho u \vec{v}\right)
  = -\grad\cdot \vec{J}_{q} + \rho\dot{q} -P \grad \cdot \vec{v} + \sum_s \vec{w_s} \vecdot \vec{f}_{s}~,
  \label{eq:internal-h}
\end{equation}
where $\dot{q}$ reflects the rate of processes that can directly
remove energy from the system (i.e., optically-thin cooling) and where
$\vec{J}_q$ is the ``heat flow'' vector (e.g., thermal conduction or radiative transfer).
Denoting the Lagrangian time derivative as  $\Dif /\Dif t$,
the flow of energy into a Lagrangian parcel is
\begin{equation}
  \DDt{q} = \dot{q} - \frac 1 \rho \grad \vecdot \vec{J}_q ~,
  \label{eq:heat-flow}
\end{equation}
which when combined with the Lagrangian derivative of specific internal energy,
\begin{equation}
  \DDt{u} = \frac{1}{\rho} \left[ \ddt{(\rho u)} + \grad \vecdot \left(\rho u \vec{v}\right)\right]~,
\end{equation}
yields
\begin{equation}
  \DDt{u} = \DDt{q} - P \DDt{}\left(\frac{1}{\rho}\right) 
  + \frac{1}{\rho} \sum_s \vec{w}_s \vecdot \vec f_{s} ~.
  \label{eq:my-first-law}
\end{equation}
When $\vec{w}_s = 0$,
Equation~(\ref{eq:my-first-law}) reduces to the familiar thermodynamic relation
\begin{equation}
  \DDt{q} = \DDt{u} + P \DDt{}\left(\frac{1}{\rho}\right),
  \label{eq:mesaiv-60}
\end{equation}
which is also Equation~(60) in \citet{Paxton2018}.

The $\vec{w}_s$ do not lead to a net transport of mass
$(\sum_s \rho_s \vec{w}_s = 0)$ or of charge $(\sum_{s} q_s n_s \vec{w_s} = 0)$. With the force density of Equation~\eqref{eq:the-force}, these
constraints mean that the final term of
Equation~\eqref{eq:my-first-law} vanishes even for non-zero
$\vec{w}_s$.  Therefore, the ``first law'' expression in
\citet{Paxton2018} need not be modified, even in this non-equilibrium
circumstance.

\subsection{Entropy and the Second Law}

The entropy is a quantity that can be defined in terms of the
macroscopic characteristics of the system.  We make the assumption
that, even though the star itself is not in equilibrium, there are
small mass elements in a local equilibrium where the local entropy is
well characterized. It is often useful to characterize heating by a
physical process like mixing in terms of entropy generation and
evolution (e.g., \citealt{Beznogov2013}), so here we derive the
relationship between entropy and the thermodynamic relations of the
previous section.
If we write the internal energy, $U$, of such an element in terms of
the independent thermodynamic basis variables $(S,V,N_s)$, then
expanding yields the thermodynamic identity
\begin{equation}
  \label{eq:thermo-id}
   \dif U = T \dif S - P \dif V + {\sum_s \mu_s \dif N_s}~,
\end{equation}
where $S$ is the entropy and $T$ is the temperature.  The sum runs
over all species present and $\mu_s$ is the chemical potential for
species $s$.
Writing these differentials as time derivatives and casting in specific (per unit mass) form gives
\begin{equation}
  T \DDt{s} = \DDt{u} +  P \DDt{}\left(\frac{1}{\rho}\right) - \sum_s \mu_s \DDt{}\left(\frac{n_s}{\rho}\right)~.
\end{equation}

From continuity (both individual species and total),
\begin{equation}
  \DDt{}\left(\frac{n_s}{\rho}\right) =
  -\frac{1}{\rho}  \grad \cdot \left(n_s \vec{w}_s \right)~,
\end{equation}
so that
\begin{equation}
  T \DDt{s} = \DDt{u} +  P \DDt{}\left(\frac{1}{\rho}\right) + \frac{1}{\rho}{\sum_s \mu_s \grad \cdot \left(n_s \vec{w}_s \right)}~.
\end{equation}
Combining with Equation~\eqref{eq:my-first-law} and manipulating
the term with the divergence, we have
\begin{equation}
  T \DDt{s} = \DDt{q} +  \frac{1}{\rho} \sum_s \vec{w}_s \vecdot \vec f_{s} +
  \frac{1}{\rho}{\sum_s \left[\grad \cdot\left( n_s \mu_s \vec{w}_s \right) -  n_s \vec{w_s} \cdot \grad \mu_s \right] }~.
\label{eq:manipulate}
\end{equation}
We can rearrange terms to write
\begin{equation}
  T \DDt{s} = \DDt{q} +  \frac{1}{\rho} \sum_s \grad \cdot\left( n_s \mu_s \vec{w}_s \right)
  + \frac{1}{\rho} \sum_s \vec{w}_s \vecdot \left(\vec f_{s} - n_s \grad \mu_s\right)   ~.
  \label{eq:Tds-final}
\end{equation}
Note that in the notation of \citet{Beznogov2013} this last term in
parentheses is $n_s \tilde{\boldsymbol{F}}_{s}$. They say this term
(their Equation~20) is the rate of specific entropy generation via
collisions.  We will see that is true in the case of constant $T$
(which they assume early on).

Note that this collisional entropy production in the final term
of Equation~\eqref{eq:Tds-final} need not always be associated with
production of ``heat'', as there is freedom in
Equation~\eqref{eq:Tds-final} for a suitably defined entropy of mixing
with $\Dif q/\Dif t = 0$. In terms of implementation, this implies
that the basis for defining the specific entropy $s$ in an equation of state (EOS) must be consistent with the composition degrees of freedom where
mixing may produce entropy if Equation~\eqref{eq:Tds-final} is to be
used. For example, an EOS that defines the compositional dependence
of entropy in terms of only the average particle mass would not be adequate for using
Equation~\eqref{eq:Tds-final} to study entropy generation by
sedimentation of multiple distinct species of particles with the same
mass.

By defining the entropy flow as
\begin{equation}
  \vec{J}_s = \frac{1}{T}\left(\vec{J}_q-\sum_s \mu_s n_s \vec{w}_s\right) ~,
\end{equation}
we can rewrite Equation~\eqref{eq:Tds-final} as an entropy balance
equation
\begin{equation}
  \rho \DDt{s} = -\grad \cdot \vec{J}_s + \sigma ~,
\end{equation}
where the entropy generation rate is
\begin{equation}
  \sigma = \frac \rho T \DDt{q} - \frac{1}{T^2} \vec{J}_{q} \cdot \grad T + \frac{1}{T} \sum_ s \vec{w}_s \vecdot \left(\vec f_{s} - n_s T \grad \frac{\mu_s}{T}\right)~.
  \label{eq:final-rate}
\end{equation}
Setting $\Dif q/\Dif t = 0$, so that one can meaningfully define a closed
system, then the constraint that each term in $\sigma$ must be
non-negative leads to the satisfaction of the second law of
thermodynamics.  In Chapter
III, \citet{dGM} discuss these points further and show several
equivalent ways of writing this expression using different definitions
of the heat flow.

\section{Implementation of Sedimentation Heating}
\label{sec:conceptual}

We now revisit the energetics of gravitational settling and demonstrate the relationship between the the \citet{Paxton2018} and
\citet{GarciaBerro2008} approaches.  This clarifies the approximation
being made in \MESA.

\subsection{Description of Approaches}

When formulating the stellar structure equations, energy conservation
is included by considering the energy flow in and out of a fluid
parcel \citep[e.g.,][Chapter 4]{Kippenhahn2012}.
In this Lagrangian picture, to understand how the energy of a
fluid parcel is changing, we account for the specific (per unit
mass) rate of energy injection into the parcel, $\epsilon$, and the
specific rate of energy flow through the boundaries,
$\partial L/\partial m$.  The specific heating rate ($\Dif q/\Dif t$)
for the parcel then satisfies
\begin{equation}
  \label{eq:local-energy}
\DDt{q} = \epsilon - \ddm{L}.
\end{equation}
Using Equation~\eqref{eq:mesaiv-60},
this is traditionally rewritten in terms of a source function
$\epsgrav$, such that
\begin{equation}
\ddm{L} = \epsilon + \epsgrav~,
\end{equation}
with
\begin{equation}
  \epsgrav \equiv -\DDt{u} - P \DDt{}\left(\frac{1}{\rho}\right)~.
  \label{eq:eps-grav}
\end{equation}

\subsubsection{Approach in \citet{GarciaBerro2008}}

\citet{GarciaBerro2008} follow this approach, with
$\epsilon = -\epsilon_\nu$ (only optically-thin neutrino cooling, no
nuclear reactions).  They restrict themselves to a composition with
two chemical elements, so that $X_{22} + X_{b} = 1$.%
\footnote{\citet{GarciaBerro2008} label their species with subscripts
  1 and 2.  We label their species ``1'' with ``22'' to indicate it is
  \neon[22]; we label quantities associated with their species ``2''
  with ``b'' (to represent the background).}  When $\Dif u/\Dif t$ is evaluated in the $(\rho, T, X_{22})$ basis, they derive their Equation~(5), which in our notation is
\begin{equation}
  \label{eq:GB08-eq5}
  \ddm{L} = -\epsnu +
  \frac{T}{\rho^2} \left(\frac{\partial P}{\partial T}\right)_{\rho, X} \DDt{\rho} - c_{V} \DDt{T} - \left(\frac{\partial u}{\partial X_{22}}\right)_{\rho,T} \DDt{X_{22}}~,
\end{equation}
where $c_V = (\partial u/\partial T)_{\rho,X}$ is the specific heat at constant volume.
We identify the effective \citet{GarciaBerro2008} \neon[22] heating term as
\begin{equation}
  \epsne \equiv -\left(\frac{\partial u}{\partial X_{22}} \right)_{\rho, T}  \DDt{{X}_{22}}~.
  \label{eq:eps22-GB08}
\end{equation}

\subsubsection{Approach in \citet{Paxton2018}}

When evaluating the total derivative of the internal energy in
Equation~\eqref{eq:eps-grav}, a common approximation is to neglect the
composition derivatives.  This is the default approach adopted in
\MESA, justified by the fact that these terms are negligible compared to the energy released by nuclear reactions
\citep[e.g.,][]{Kipp65}.  When $(\rho,T,\{X_i\})$ are the structure
variables, the usual form of \epsgrav\ in \MESA\ is given in
Equation~(12) of \citet{Paxton2011} or Equation~(65) of \citet{Paxton2018}:
\begin{equation}
  -\epsgrav = c_P T \left[ \left( 1 - \nabla_{\rm ad} \chi_T \right) \DDt{ \ln T} - \nabla_{\rm ad} \chi_\rho \DDt{ \ln \rho}  \right]~.
  \label{eq:mesa-eps-grav}
\end{equation}
The thermodynamic derivatives have their usual definitions:
$c_P = (\partial e/\partial T)_{P,X}$,
$\chi_T = (\partial\ln P/\partial\ln T)_{\rho,X}$,
$\chi_\rho = (\partial\ln P/\partial\ln \rho)_{T,X}$, and
$\nabla_{\rm ad}= (\partial \ln T/\partial \ln P)_{s, X}$.
Equation~\eqref{eq:mesa-eps-grav} is identical (after using
thermodynamic and mathematical identities) to the non-$\epsilon_{22}$
terms in Equation~\eqref{eq:GB08-eq5}.

Evaluating the neglected composition derivative terms can be difficult
in a code like \MESA\ when accounting for a mixture of more than a few
elements.
The model evolves the set of mass fractions specified by the
choice of nuclear network.  However, this structural basis
$(\rho, T, \{X_i\})$ usually does not match the basis of the EOS.  For
the $(\rho, T)$-basis EOSes relevant for WDs in \MESA, the composition
basis for HELM is $(\Abar, \Zbar)$ while the composition basis for PC
is all isotopes with mass fractions above some threshold (default:
$10^{-3}$).  This makes it challenging (or impossible) to evaluate all the
relevant partial derivatives with respect to composition within \MESA.

Therefore, instead of attempting to fully evaluate $\epsgrav$, the
energetic effects of \neon[22] sedimentation are incorporated into
\MESA\ via the inclusion of an additional heating term specific to
this isotope.  This follows the approach taken by \citet{Bildsten2001}
and \citet{Deloye2002} who evaluate the net power generated by
\neon[22] sedimentation in the trace limit.  The specific rate at
which energy is deposited is then
\begin{equation}
\begin{aligned}
\epsilon_{\rm 22} 
= \frac{ |F| v_{22}} {(A m_{\rm p})/X_{22}}
\label{eq:ne22_heat}
= \left( 22 m_{\rm p} g - 10 e E  \right) \frac{X_{22} v_{22}}{22 m_{\rm p}}.
\end{aligned}
\end{equation}
The $^{22}{\rm Ne}$ diffusion velocity ($w_{22}$) and electric field
($E$) are calculated in the diffusion routine and then used to
evaluate the above heating term.
This expression assumes the only isotope settling is \neon[22] and
assumes it is a trace ($X_{22} \ll 1$).
By assuming only two isotopes, \citet{GarciaBerro2008}
also makes the assumption that only \neon[22] sediments, though their
expression does not explicitly make the trace assumption.\footnote{Since
  \neon[22] mass fractions are typically of order the initial
  metallicity, the trace limit is initially a good approximation.
  It does become worse as \neon[22] centrally concentrates (see e.g.,
  Figure~\ref{fig:comp} where the mass fraction reaches $\approx 0.2$).
}

\subsection{Unification of the two approaches}

The heating terms in Equation~\eqref{eq:eps22-GB08} and
Equation~\eqref{eq:ne22_heat} appear different.  In particular, the
\MESA\ approach indicates that heating occurs anywhere that the
\neon[22] has a finite abundance and a non-zero diffusion velocity,
while the
\citet{GarciaBerro2008} approach only indicates heating only where the
local \neon[22] abundance is changing.
White dwarf cooling calculations with \MESA\ were compared with the
results from \citet{GarciaBerro2008} and found to agree \citep[see
Section 3.5, Figure 18 in][]{Paxton2018}.  But if one inspects the
\textit{local} profiles of the \neon[22] heating in \MESA\ and in
Figure~4 of \citet{GarciaBerro2008}, the two approaches appear
different.

\subsubsection{Approach of \citet{GarciaBerro2008}}

Let us evaluate the total heating rate of sedimentation over the whole star,
\begin{equation}
  \dot{E}_{22} = \int \epsne\ dm = -\int dm \left(\frac{\partial u}{\partial X_{22}} \right)_{\rho, T}  \DDt{{X}_{22}}~.
\end{equation}
Since there are no nuclear reactions, composition changes are entirely
due to gravitational settling, so that
\begin{equation}
   \DDt{{X}_{22}} = - \frac{1}{\rho} \grad \cdot \left(\rho X_{22} \vec{w}_{22} \right)~.
\end{equation}
As \citet{GarciaBerro2008} indicate, in a cool WD the electronic
contributions to the internal energy dominate.  In the fully
degenerate limit, 
\begin{equation}
  \left(\frac{\partial u}{\partial X_{22}} \right)_{\rho, T} =
  \left(\frac{\partial u}{\partial \Ye} \right)_{\rho, T}
  \left(\frac{\partial \Ye}{\partial X_{22}} \right) \approx
  \frac{\mue}{\amu} \left(\frac{10}{22} - \frac{Z_b}{A_b} \right)= 
-\frac{\mue}{22 \amu} ~,
\end{equation}
where $\amu$ is the atomic mass unit, $\mue$ is the electron chemical potential, and we have assumed the background species has $Z_b/A_b$ =
1/2.  Therefore, we have
\begin{equation}
  \begin{split}
    \dot{E}_{22} & = -\frac{1}{22 \amu} \int \frac{\mue}{\rho} \grad \cdot \left(\rho X_{22} \vec{w}_{22}\right) dm \\
    & = -\frac{1}{22 \amu} \int \frac{1}{\rho} \left[\grad \cdot \left(\mue \rho X_{22} \vec{w}_{22}\right) - \rho X_{22} \vec{w}_{22} \cdot \grad \mue \right] dm \\
  &=
-\frac{1}{22 \amu} \left[\int \ddm{}\left(4 \pi r^2 \mue \rho X_{22} \vec{w}_{22}\right) dm -
  \int X_{22} \vec{w}_{22} \cdot \grad \mue\ dm \right] \\
&=  \frac{1}{22 \amu} \int X_{22} \vec{w}_{22} \cdot \grad \mue\ dm ~.
\end{split}
\end{equation}
The manipulation of the divergence is analogous to that in
Equation~\eqref{eq:manipulate}.  The first integral on the third line
vanishes since the term in parentheses vanishes at the center and the
surface.  Diffusion changes the local entropy both by transporting
entropy and locally generating entropy. Here we see that a term
similar to local entropy generation is closely related to the total
integrated change in internal energy, while the integrated transport
term has no net effect.

The thermal conduction timescale across the WD core is much less than the
evolutionary timescale.  This has the consequence that the transport
term can be neglected.  To get the overall cooling right, it does not
matter exactly how the heating is deposited, so long as the correct total
amount of energy is deposited.  Thus, an equivalent
expression for \epsne\ is given by the part of the integrand that
does not vanish,
\begin{equation}
  \epsne = \frac{1}{22 \amu} X_{22} \vec{w}_{22} \cdot \grad \mue~.
  \label{eq:epsne-gb08-alt}
\end{equation}

\subsubsection{Approach in \citet{Paxton2018}}
\label{sec:unifying_paxton}

Give the final term in Equation~\eqref{eq:Tds-final} the
name
\begin{equation}
    \epsdiff = \frac{1}{\rho} \sum_s \vec{w}_s \vecdot \left(\vec f_{s} - n_s \grad \mu_s\right)   ~.
\end{equation}
As shown in Equation~\eqref{eq:final-rate}, under the assumption of
constant $T$, this is proportional to the local entropy generation rate due to
diffusion.  Using the definition of $\vec{f}_s$ we write
\begin{equation}
  \epsdiff = \frac{1}{\rho} \sum_s n_s \vec{w}_s \vecdot \left(m_s \vec{g} + q_s \vec{E} - \grad \mu_s\right)   ~.
  \label{eq:epsdiff-expanded}
\end{equation}

To satisfy charge neutrality, the
electron flux must be similar to the largest ion fluxes $| n_e
\vec{w}_e | \sim | n_i \vec{w}_i|$. In the strongly degenerate
limit, the electrons are responsible for maintaining hydrostatic
equilibrium, so that the term in parentheses in
Equation~\eqref{eq:epsdiff-expanded} must be very close to zero for
electrons.
Hence the sum in Equation~\eqref{eq:epsdiff-expanded} can be
restricted to run only over the ion species, since the entropy
generation for the electrons is negligible near equilibrium, while the
deviations from equilibrium can be much larger for an ion such as
\neon[22] where $\grad \mu_i$ is negligible and the electric and
gravitational forces do not cancel.

The electric field estimate of Equation~\eqref{eq:e-field} is
\begin{equation}
  \vec{E} = -\frac{\Abar \amu }{\Zbar e} \vec{g}~.
\end{equation}
where $\Abar$ and $\Zbar$ are the mean ion mass number and charge
number, respectively.
Dropping the ion chemical potentials and using this estimate
of the electric field in Equation~\eqref{eq:epsdiff-expanded} gives
\begin{equation}
\epsdiff = \frac{1}{\rho} \sum_{\text{ions}} \vec{w}_s \vecdot  \vec{g} \left(\rho_s - n_s Z_s \frac{\Abar \amu}{\Zbar}\right)~.
\end{equation}
For ions, the mass fraction is defined as $X_s = \rho_s/\rho$ and 
$n_s = \rho_s / (A_s \amu)$, so
\begin{equation}
\epsdiff = \vec{g} \vecdot \sum_{\text{ions}} \vec{w}_s X_s \left(1 - \frac{Z_s/A_s}{\Zbar/\Abar}\right) ~.
\end{equation}
This term only appears if the charge-to-mass ratio of the species
being transported is different than the background.  In the limit in
which all other isotopes have $Z_s/A_s = 1/2$ and $ \Zbar/\Abar = 1/2$ (trace limit),
the only non-zero term is that for \neon[22].  That is,
\begin{equation}
  \epsdiff = \epsilon_{22} = 2 g w_{22} X_{22} / 22~,
  \label{eq:eps22-mesa}
\end{equation}
which is equivalent to Equation~(16) in \citet{Paxton2018}.

\subsection{Demonstration of Equivalence}
Alternatively, taking the sum over all
species (including electrons) in Equation~\eqref{eq:epsdiff-expanded},
the gravitational and electric force terms vanish due to the lack of
net flow of mass and charge.  The term with the electron chemical
potential gradient remains, implying
\begin{equation}
  \epsdiff = -\frac{n_e}{\rho} \vec{w}_e \cdot \grad \mu_e = -\frac{\Zbar}{\Abar \amu} \vec{w_e} \cdot \grad \mu_e ~.
  \label{eq:epsdiff-electron}
\end{equation}
The constraints of no net transport of mass or charge 
also imply that the electron velocity is related to the ion velocity.   In the context of two species, the mass constraint is
\begin{equation}
  X_{22} \vec{w}_{22} + (1-X_{22})\vec{w}_{b} = 0  
\end{equation}
and the charge constraint is
\begin{equation}
  \frac{\Zbar}{\Abar} \vec{w}_e = X_{22} \frac{Z_{22}}{A_{22}} \vec{w}_{22} + (1-X_{22})  \frac{Z_{b}}{A_{b}} \vec{w}_{b} ~.
\end{equation}
Combining these, and assuming a background with $Z_b/A_b = 1/2$, we have
\begin{equation}
  \frac{\Zbar}{\Abar} \vec{w}_e =
  \left(\frac{10}{22} - \frac{Z_b}{A_b} \right) X_{22} \vec{w}_{22}
  = -\frac{X_{22}}{22} \vec{w}_{22} ~.
\end{equation}
This demonstrates that Equation~\eqref{eq:epsne-gb08-alt} and
Equation~\eqref{eq:eps22-mesa} are equivalent.

This clarifies that the  \MESA\ approach retains
only the local entropy generation term in
Equation~\eqref{eq:local-energy} and shows why \citet{GarciaBerro2008}
and \MESA\ agree on the net effect of sedimentation on WD cooling
despite the apparently different approaches.

\section{Charge Stratification}
\label{sec:charge}

In this section, we 
tentatively rule out any physical process that relies on
stratification of elements by charge as a source of $\sim$10~Gyr
cooling delays.
In strongly-coupled Coulomb mixtures, ions of higher-than-average
charge also sediment, even with equal charge-to-mass ratio
\citep{Chang2010,Beznogov2013}. The forces associated with this
tendency toward charge stratification can modify
diffusion velocities of elements responsible for sedimentation
heating, and movement of ions through the chemical potential gradient
driving this charge separation yields an additional heating
term (Equation~\ref{eq:epsdiff-expanded}). However, we show here that this term is small compared to
single-particle $^{22}$Ne sedimentation heating in WD interiors.

\begin{figure}
\begin{center}
\includegraphics[width=\apjcolwidth]{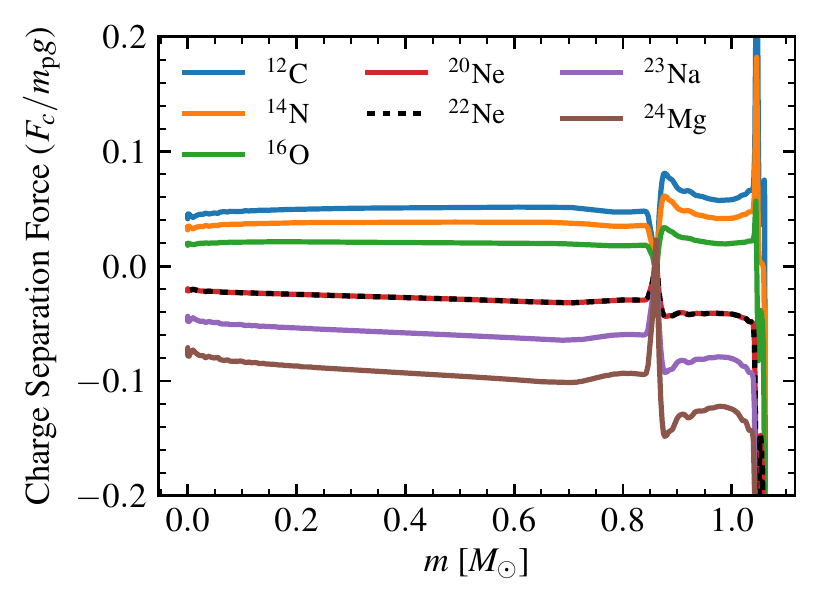}
\caption{Example of the charge separation forces in a $1.06~M_\odot$
  O/Ne WD. This model has cooled to a luminosity of
  $10^{-2}~L_\odot$ and will soon begin core crystallization.}
\label{fig:separation}
\end{center}
\end{figure}

Following \cite{Beznogov2013}, for species $j$ of charge $Z_j$, the Coulomb
interactions with other ions give rise to a chemical potential of the
form
\begin{equation}
\mu_j^{(C)} = -0.9 \frac{Z_j^{5/3}e^2}{a_{\rm e}}~,
\end{equation}
and therefore the average Coulomb chemical potential over all ions in
the plasma is
\begin{equation}
\langle \mu^{(C)} \rangle  = -0.9 \Gamma k_{\rm B} T~.
\end{equation}
The charge separation force for element $j$ relative to the background
of all ions is then (cf.\ equation~(14) of \citealt{Beznogov2013})
\begin{equation}
F_{c,j} \equiv
\frac{1}{\langle Z_{\rm i} \rangle}\left[ Z_j \nabla \langle \mu^{(C)} \rangle
  - \langle Z_{\rm i} \rangle \nabla \mu_j^{(C)} \right] 
\approx \left(\frac{Z_j}{\langle Z_{\rm i} \rangle} 
  - \frac{Z_j^{5/3}}{\langle Z_{\rm i}^{5/3} \rangle} \right) \nabla \langle \mu^{(C)} \rangle ~,
\end{equation}
where we have assumed that $\nabla \langle Z_{\rm i}^{5/3} \rangle$ is
negligible for the final step.
Figure~\ref{fig:separation} shows this estimated force relative to the
local value of $m_{\rm p} g$ for each species of ion throughout the
interior profile of a $1.06~M_\odot$ \MESA\ WD model with an
O/Ne dominated core. While these forces may be enough to drive
small amounts of diffusion, they are much smaller than the
gravitational sedimentation forces ($F_g \geq m_{\rm p}g$) that act
on neutron-rich isotopes such as $^{22}$Ne and $^{23}$Na. Diffusion
velocities are linearly proportional to the forces driving them $v
\propto F$, and heating associated with these velocities scales as
$\epsilon \propto vF \propto F^2$. Diffusion related to charge
separation will therefore lead to negligible heating compared to
gravitational sedimentation, so we are justified in ignoring this term
in the models presented in this work.

Figure~\ref{fig:separation} also allows an estimate of the total
amount of energy that could be released associated with 
any other mixing that might lead to rearrangement of charges in
this chemical potential (e.g.\ phase separation induced by
crystallization) using Equation~\eqref{eq:delay}.
For the most abundant elements (O~and Ne in this \MESA\ model),
complete charge stratification would lead to a delay of approximately 
5~Gyr. This is somewhat smaller than the energy and cooling delay
available from $^{22}$Ne sedimentation. 
The ratio of the total energies from both of these
sources is approximately
$E_g/E_c \approx (2 X_{\rm 22})(F_g/F_c) \approx 200 X_{22}$.
So the total time-delay available from charge separation is comparable
to that available from complete $^{22}$Ne when 
metallicity is $Z\approx 0.005$, but WDs descended from metal-rich
stars have a potential sedimentation energy reservoir several times
larger.

We also note that this energy estimate predicts that cooling delays
comparable to those observed for the Q branch could only result from
phase separation if chemical separation of the dominant elements (C/O
or O/Ne) is nearly complete, releasing all of the energy associated
with these chemical potential differences. In fact, phase separation
calculations do not predict complete separation, and hence cooling
delays associated with phase separation are typically on the order of
1~Gyr \citep{Segretain94,Chabrier00,Althaus12}, insufficient
to explain the full extent of the cooling anomaly on the Q branch.

\end{CJK*}

\clearpage
\bibliographystyle{yahapj}
\bibliography{paper}

\end{document}